\documentclass[nofootinbib,aip,amsmath,amssymb,reprint,twocolumn,10pt]{revtex4-2}

\usepackage{graphicx}
\usepackage{dcolumn}
\usepackage{bm}
\usepackage[utf8]{inputenc}
\usepackage[T1]{fontenc}
\usepackage{mathptmx}
\usepackage{verbatim}
\usepackage{fdsymbol}
\usepackage{graphicx}
\usepackage{sidecap}
\usepackage{placeins}
\usepackage{xcolor}
\usepackage{amsfonts,amsthm,amsmath,amssymb,amscd,graphicx}
\usepackage{mathtools}
\usepackage{hyperref}
\usepackage{amsmath}
\usepackage{amssymb}
\usepackage{bbold}
\usepackage{float}
\usepackage{hyperref}

\begin{document}

\title{Building blocks of rhythmic neural circuits  through cellular and synaptic dynamics}
\title{Complementing cellular and synaptic dynamics for building blocks of rhythmic neural circuits}
\title{Pairing cellular with synaptic dynamics to build rhythmic neural blocks}
\title{Blending cellular and synaptic dynamics into building blocks of rhythmic neural circuits}

\title{Pairing cellular and synaptic dynamics into building blocks of rhythmic neural circuits}

\author{James Scully}
\email{jscully2@student.gsu.edu}
\affiliation{Neuroscience Institute,  Georgia State University, 
100 Piedmont Ave., Atlanta, GA 30303, USA.}
\author{Jassem Nabeel Hassan Bourahmah}
\email{jbourahmah1@student.gsu.edu>}
\affiliation{Neuroscience Institute,  Georgia State University, 
100 Piedmont Ave., Atlanta, GA 30303, USA.}
\author{David Bloom}
\email{dbloom2@student.gsu.edu}
\affiliation{Neuroscience Institute,  Georgia State University, 
100 Piedmont Ave., Atlanta, GA 30303, USA.}

\author{Andrey L. Shilnikov}
\email{ashilnikov@gsu.edu}
\affiliation{Neuroscience Institute and Department of Mathematics \& Statistics, Georgia State University, \\
100 Piedmont Ave., Atlanta, GA 30303, USA.}

\date{\today}

\begin{abstract}

The purpose of this paper is trifold -- to serve as an instructive resource and a reference catalog for biologically plausible modeling with i) conductance-based models, coupled with ii) strength-varying slow synapse models, culminating in iii) two canonical pair-wise rhythm-generating networks. We document the properties of basic network components: cell models and synaptic models, which are prerequisites for proper network assembly. Using the slow-fast decomposition we present a detailed analysis of the cellular dynamics including a discussion of the most relevant bifurcations. Several approaches to model synaptic coupling are also discussed, and a new logistic model of slow synapses is introduced. Finally, we describe and examine two types of bicellular rhythm-generating networks: i) half-center oscillators ii) excitatory-inhibitory pairs and elucidate a key principle -- the network hysteresis underlying the stable onset of emergent slow bursting in these neural building blocks. These two cell networks are a basis for more complicated neural circuits of rhythmogensis and feature in our models of swim central pattern generators.
        
\end{abstract}

\maketitle

\section{Introduction}

The small neural circuits, called central pattern generators (CPG) \cite{CPG, Marder-Calabrese-96, KatzHooper07}, that drive simple locomotion patterns in invertebrates contain surprising complexity. Biologically plausible modeling of locomotion presents an opportunity to address subtleties in the interpretation of biological results, and to provide analysis tools that will be useful to conductance-based modelers outside of the invertebrate CPG niche. There is a large variety of CPG networks reported in the literature. Some CPGs are composed of endogenous bursters driven by pacemakers while other CPGs generate stably slow oscillations through network-level mechanisms. The best-known example of such a circuit with the pacemaker is the pyloric CPG which produces a bursting rhythm to control striated muscles that dilate and constrict the pyloric region of the lobster stomach. \cite{selverston1976stomatogastric,stein1999neurons,selverston2013model,Marder1994752,prinz2003functional,prinz2004similar, M09, M12}. 

This paper has evolved out of collaborative efforts to model the undulating swim rhythms of two sea slugs, {\it Melibe leonina}\cite{NW02, WNT02, TW05} and {\it Dendronotus iris}\cite{SNLK11} . The CPG circuits underlying their behaviors have been studied extensively in both species \cite{NSLGK012,sck2014, sakurai2015phylogenetic, sakurai2016central}. All neurons in the CPGs have been identified, and their synaptic connections have been determined with careful pairwise electrophysiological recordings \cite{sakurai2017artificial,sakurai2019command,sakurai2022bursting}. The swim CPGs in these sea slugs do not include endogenously bursting pacemakers. Hence, each circuit should be viewed as a whole rather than by looking at specific pacemaker cells. The working hypothesis is that the given CPG circuits function at controlled oscillatory states emerge largely through the synergetic interactions among the coupled components with similar dynamic and nonlinear properties. This is the main driver and the starting point of our computational study, which is focused on identifying what cellular and synaptic qualities can warrant robust generation of slow oscillations in two specific bicellular blocks, which happened to be symmetrically built into the larger CPG circuits in the given sea slugs.  
   
The organization of this paper serves to develop basic principles of stable bursting emerging in neural building blocks with properly functioning and matching cellular and synaptic components. We will begin with the introduction and slow-fast bifurcation analysis of the conductance-based cellular model to describe the basic properties of the swim CPG interneurons in Section~\ref{sec:2}. The following Section~\ref{sec:3} is dedicated to modeling slow synaptic dynamics and examines how the strength of synaptic coupling may changes with the spike frequency variations in the neuron model. In the Section ~\ref{sec:4} based on the proposed concept of a {\em network hysteresis}, we will introduce two types of building blocks for larger neural circuits, including the model of a half-center oscillator (HCO) in the Section~\ref{sec:5}, which a pair of neurons reciprocally coupled with slow inhibitory synapses and elucidate basic principles underlying the stable onset of emergent slow bursting in neural circuits. The following Section~\ref{sec:6} generalizes the principles for network-level bursting emerging in a bicellular excitatory-inhibitory module. The concluding Section~\ref{sec:7} discusses module organizations of the swim CPG circuits of sea slugs {\it Melibe leonina} and {\it Dendronotus iris} and how specifically they can be assembled with the examined building blocks.
 	
\section{Conductance-based model of swim CPG interneurons} \label{sec:2}

Our objective in choosing a neuron model was its biological plausibility, even though cellular currents in swim CPG interneurons of the sea slugs {\it Melibe leonina} and {\it Dendronotus iris} have not been identified or specified.

So, as our starting point for a conductance-based model to describe the swim CPG interneurons we picked the Plant model \cite{plant75,plant76,plant81} derived from the Hodgkin-Huxley (HH) formalism. The model was originally introduced to examine endogenously {\em parabolic} bursting observed and recorded in the neuron R15 located in the abdominal ganglion of the gastropod mollusk (sea slug) {\it Aplysia californica} \cite{RP1985,LL88}. The bursting neuron R15 is hypothesized to be involved in the Aplysia egg-laying process. Our use of the Plant model was justified by our initial, and perhaps contentious observation that in experimental studies the {\it Melibe} swim CPG also produced parabolic bursting, for which the Plant model was notorious. Following the formal classification~\cite{Rinzel1985, Rinzel87b}, the spike frequency  in parabolic bursting is maximized in the middle of a spike train and noticeably decays at the end points, thereby resulting in a bent shape of the spike frequency distribution~\cite{EK1986,Rinzel1995a}. An accurate mathematical analysis of the Plant model was initially done in Refs.~\cite{RL86,RL87} using slow-fast dissection, while a variety of dynamical properties of the R15-neuron models were examined later in following publications, see  Refs.~\cite{CCB91, bertran1993, Canavier1993,  BCCBB95, butera98, R15, Alacam2015} and references therein. One of the features of the Plant model is that its dynamical variables can be separated into two subsystems with two disparate time-scales: (i) the {\em fast} subsystem for fast spike generation, which was borrowed from the original Hodgkin-Huxley model, and (ii) the {\em slow} subsystem to modulate the fast spiking activity and regulate termination and recovery which are necessary for robust endogenous bursting, as was experimentally observed in the R15 neurons. 
           
The swim CPG interneurons in {\it Melibe leonina} and {\it Dendronotus iris} differ from the R15cells in several respects, which should be properly addressed. First of all, swim interneurons have not been observed to burst endogenously. Moreover, these interneurons are not latent bursters either, as they do not burst even when perturbed with constant external currents \cite{Alacam2015}. Neurophysiological experiments on the swim CPGs\cite{SNLK11, NSLGK012,sck2014, sakurai2015phylogenetic, sakurai2016central,sakurai2017artificial,sakurai2019command,sakurai2022bursting} have demonstrated that their interneurons are either quiescent at most times, or become tonic-spiking during swim episodes when receiving excitatory drives from sensory cells. This strongly suggests that the slow bursting (of period ranging from 2 through 14 sec, resp., in juvenile and grown animals) observed in the experimental studies on the swim CPGs in the sea slugs is indeed a {\em network-level} dynamical phenomenon due to nonlinear interactions between the interneurons orchestrated by complex coordination of fast and slow currents, including synaptic ones. In our modeling efforts, it is important to address the observed activity of the CPG interneurons in normal function and under perturbation, as well as describe the realistic timescales of various synapses in order to explore the network-level rhythmogenesis in such circuits.

\subsection{Short description of the adapted Plant model}

\begin{figure}[!t]
\centering
\includegraphics[width=.45\textwidth]{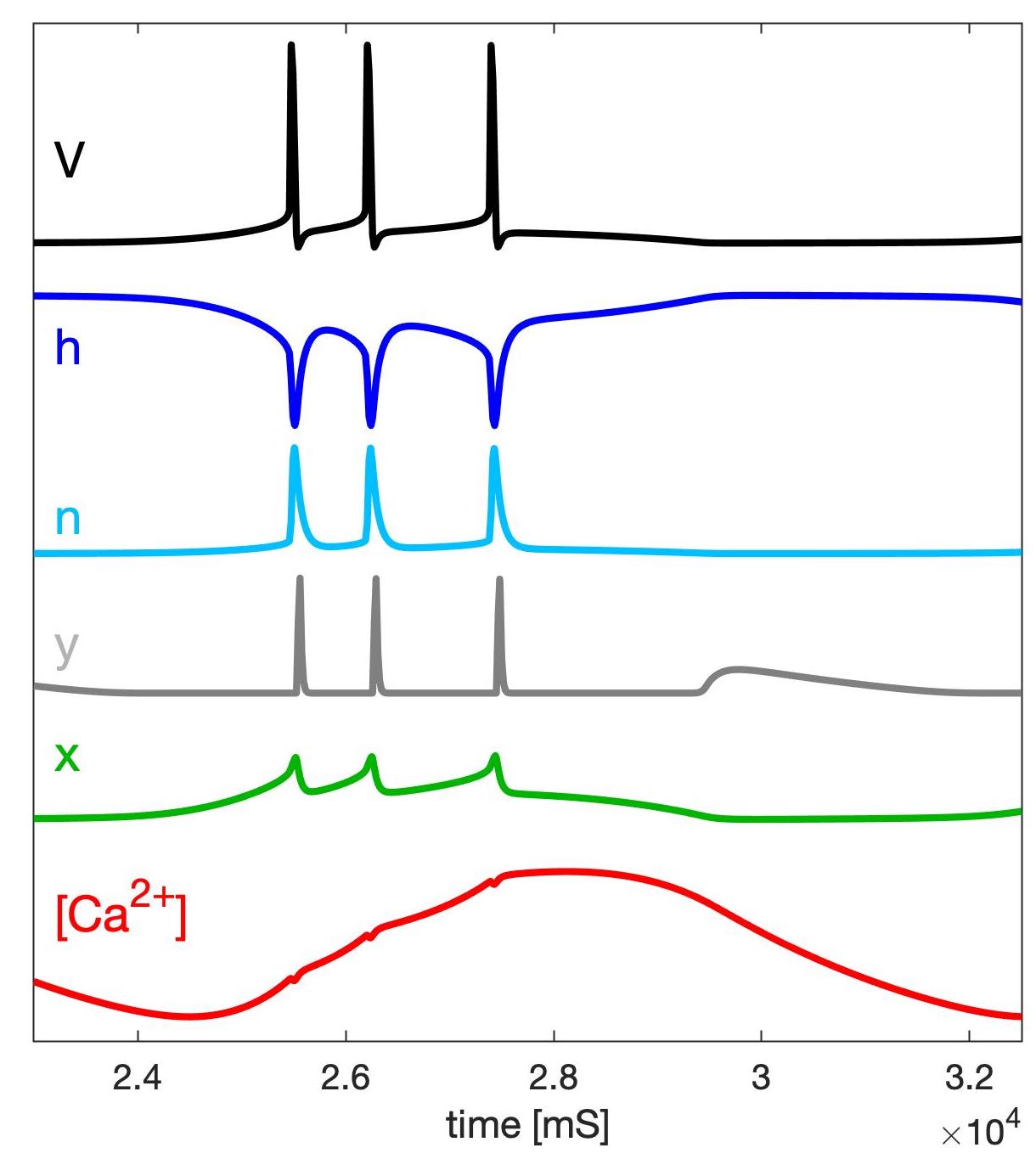}
\caption{Time scales of the fast variables of the Plant burster: membrane voltage $V$, and gating probabilities $h$, $n$ and $y$ (within the [0--1] range) for the  inward $Na^{+}$ sodium and $\rm Ca^{2+}$ calcium, $K^+$ potassium and $h$- currents , resp., compared against the slow  variables, $x$ and $[\rm Ca^{2+}]$, introduced in both calcium-coupled currents.               
 }
\label{fig:currents}
\end{figure} 

The original Plant model includes the following fast currents: the inward sodium and calcium ($I_{I}$), the outward potassium ($I_K$), necessary for the spike generation, along with the generic ohmic leak ($I_{leak}$) current. The gradual spike frequency adaptation and post-inhibitory rebound in the model are due to the slow dynamics of two currents: the TTX-resistant inward sodium and calcium current ($I_{\rm T}$) and outward calcium-sensitive potassium current ($I_{KCa}$). 
\begin{gather}
C_{m} {V}^\prime= - I_{I} - I_{K} - I_{leak} - I_{h} - I_{T} - I_{KCa}, \label{eq:voltage} ~\\ 	
h^\prime = \frac{h_{\infty}(V)-h}{ {\tau_{h}(V)} }, \qquad 
n^\prime = \frac{n_{\infty}(V)-n}{\tau_{n}(V)}, \label{eq:hngating} \\
y^\prime= \frac{1}{2} \left [ \frac{1}{1+e^{10(V-50)}}-y \right ]/ \left [ 7.1+\frac{10.4}{1+e^{(V+68)/2.2}} \right ] \label{eq:hgating}
\end{gather}
with the membrane capacitance $C_m=1$; here the dynamic variables are the membrane voltage $V(t)$, the gating probabilities $h(t)$, $n(t)$, $y(t)$ and $x(t)$ as well as the calcium concentration $[{\rm Ca}(t)]$.  This ODE system describing the R15 Plant burster includes a fast subsystem to generate repetitive tonic-spiking or quiescent activity, depending on the level of drive from its slow subsystem. This spike-generating machinery includes the following fast currents: sodium $I_{Na}$, potassium $I_K$, and additionally, leak $I_{leak} = g_{L} (V-E_{L})$, and a rapidly depolarizing $h$-current, which are given, respectively, by the following ODE equations:
\begin{align}
I_{I} &= g_{I}\,h\, m^{3}_{\infty}(V)\,(V-E_{I}), \label{eq:icurrent}\\ 
I_{K} &= g_{K} \, n^{4}\, (V-E_{K}), \label{eq:kcurrent} \\ 
I_{h} &= g_{h}\,\frac{y\,(V-E_{h}) }{\left (1+e^{-(V-63)/7.8} \right )^3 } \label{eq:hcurrent}.
\end{align}
Here, the activation of the inward sodium current is assumed instantaneous and therefore described by an analytical (sigmoidal) equation $m_{\infty}^3(V)$, rather than a corresponding fast ODE. The fourth current describes a depolarizing $h$-current that activates as the voltage drops down below $-50$mV, see Eq.~(\ref{eq:hgating}) above. 

Recurrent alternation between fast spike trains and slow quiescent episodes in the endogenous Plant burster is reciprocally modulated by the slow inward TTX-resistant $\rm Na^{+}$-$\rm Ca^{2+}$ current and slow outward $\rm Ca^{2+}$ activated $K^+$ given, respectively, by 
\begin{align} 
I_{T} &= g_{T} x (V-E_{I}), \label{eq:ttxcurrent} \\
I_{KCa} &= g_{KCa}\frac{[\rm Ca]}{0.5+[\rm Ca]}(V-E_{K}) \label{eq:cacurrent} 
\end{align}
with two dynamical variables: the calcium concentration $[\rm Ca(t)]$ and a voltage gated probability $x(t)$, which are governed by following coupled {\em slow} system:  
\begin{align}
x^\prime &= \frac{1}{\tau_x}\left [ \frac{1}{1+e^{-0.15(V+50- \Delta_{x})}}-x \right ], &\tau_x &\gg 1, \label{SlowSub1} ~\\
[\rm Ca]^\prime  &=  \rho \left ( K_{c} x (E_{Ca}-V + \Delta_{\rm Ca})-[\rm Ca] \right  ), &\rho &\ll 1,  \label{SlowSub2}  
\end{align}
where $\Delta_x$ and $\Delta_{\rm Ca}$ are new bifurcation parameters introduced to control slow dynamics of the neuron model; we will discuss their action below. The reader can find the in-detail description of all fixed biophysical parameters and their specific values in the Appendix ~\hyperref[sec:appendix1]{I} below. 

The time scales of the state variables: $V(t)$, probabilities $h(t)$, $n(t)$, $y(t)$ and $x(t)$ and calcium concentration $[\rm Ca](t)$ that gate the indicated cellular currents in the Plant model are presented in Fig.~\ref{fig:currents}. Note that depending on the chosen value of the time constant $\tau_x$, the average rate of change of the gating $x$-variable can be altered: it  can be visually as fast (at $\tau_x=100$ sec$^{-1}$) as that of the fast subsystem, or can be slowed down to match the rate of change of the $[\rm Ca]$-variable at larger values such as $\tau_x=273$ sec$^{-1}$ also used in this study. Nonetheless, the $x$-dynamics is included in the slow subsystem for a few reasons. First, for the sake of historical continuity, $x(t)$ was treated as a slow variable in the original slow-fast dissection analysis of this model proposed in Ref.~\cite{RL87}. Second, the $x$-dynamics does not contribute to spike generation, which is solely due to the fast sodium-calcium and potassium currents. Instead its reciprocal nonlinear interaction with the [\rm Ca]-dynamics is the key factor that provides the Plant model with a slow hysteresis necessary for the onset of endogenous bursting composed of alternating fast spike trains and quiescent episodes. 

Following Refs.~\cite{Alacam2015,Shilnikov2004a,Shilnikov2012}, we introduced two additional bifurcation parameters $\Delta_{\rm Ca}$ and $\Delta_x$ (measured in mV) in the slow subsystem~(\ref{SlowSub1})-(\ref{SlowSub2}) of the Plant model to calibrate its slow dynamics/kinetics, mainly to prevent it bursting. Note that variations of $\Delta_{\rm Ca}$ and $\Delta_x$ do not affect temporal characteristics of fast dynamics of the neuron model. Here, the $\Delta_x$-parameter represents a deviation from the voltage value $-50$mV at which the TTX-resistant $\rm Na^{+}$-$\rm Ca^{2+}$ current is half-activated, see   Eq.~(\ref{SlowSub1}) and the corresponding activating function $x_{\infty}(V)=1/ \left (1+e^{-0.15(V+50-\Delta_{x})} \right )=1/2$. The second parameter $\Delta_{\rm Ca}$ introduced in Eq.~(\ref{SlowSub2}) shifts the calcium reversal potential from a hypothetically high value $+140$mV. In fact, variations of $\Delta_{\rm Ca}$ act similarly in some ways to those of an externally applied current, without having significant effect on the intrinsic spiking dynamics, but changing inter-spike intervals, and thus the spike frequency of the neuron model.   

\begin{figure}[!t]
   \centering
   \includegraphics[width=.5\textwidth]{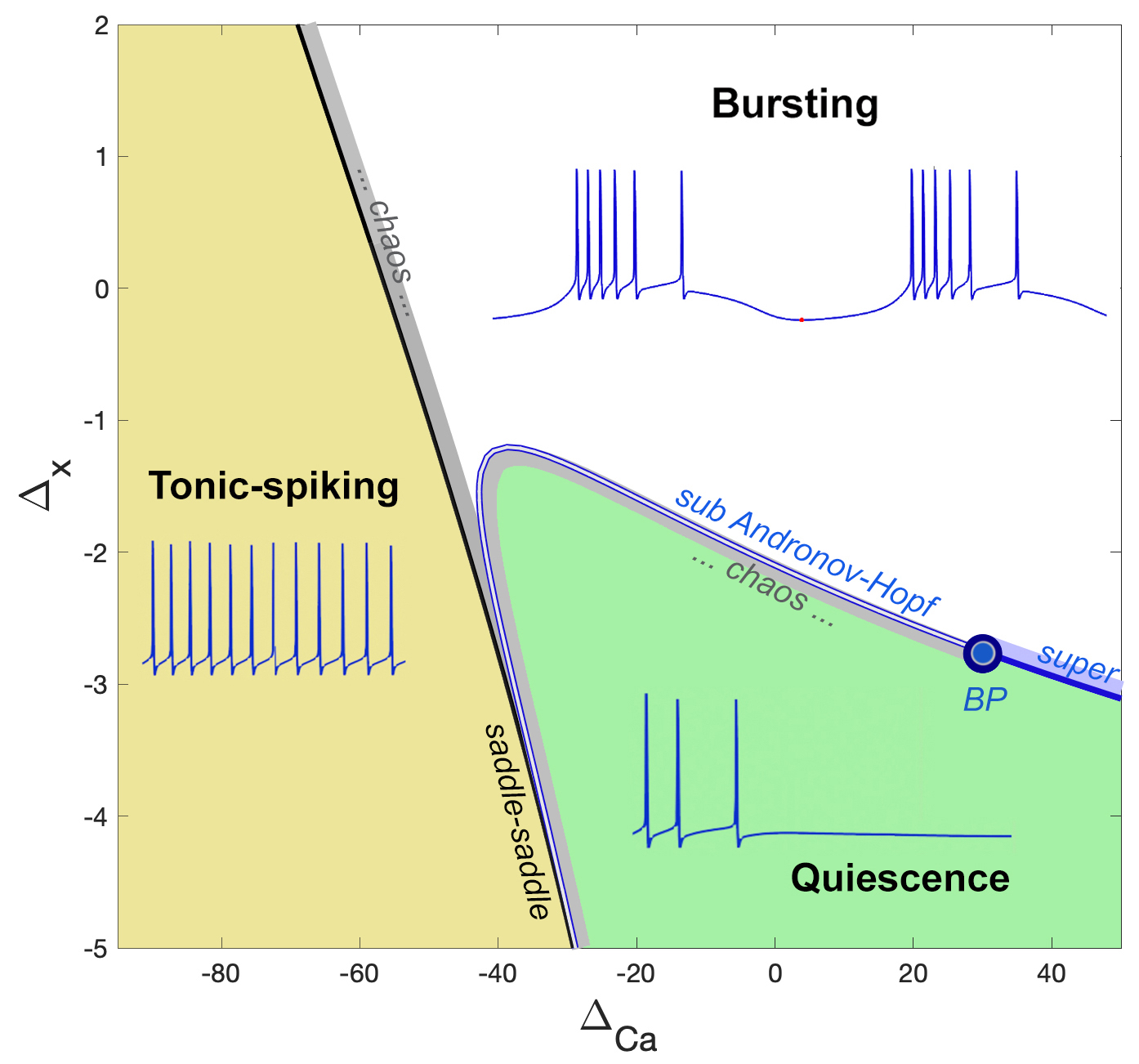}
   \caption{($\Delta_{\rm Ca}, \Delta_x$)-bifurcation diagram of the adapted Plant model with the three regions corresponding to tonic-spiking, bursting, and quiescent activity. A (horizontal) transition route from tonic-spiking to bursting activity begins with a saddle-saddle bifurcation and is followed by an adjacent narrow (grey) region of transitional chaos. A (vertical) route from quiescence to bursting passes through the Andronov-Hopf bifurcation curve with the codimension-two Bautin  point (BP) on it: on the left from the BP-point the bifurcation is a sub-critical one causing a onset of chaotic bursting within a transition layer (grey region), while on the right it is a super-critical one giving rise to small sub-threshold oscillations (in a light-blue strip) morphing into large bursting. Below the level $\Delta_x=-3.5$mV the neuron model demonstrates tonic-spiking activity or quiescence only.
   }
   \label{fig:bf_diagram}
\end{figure} 

A bifurcation diagram of the model with these parameters is presented in Fig.~\ref{fig:bf_diagram}: one can see that it is partitioned into three major regions of activity: tonic-spiking, bursting and hyperpolarized quiescent, and their borderlines in the $\left ( \Delta_{\rm Ca}, \, \Delta_x \right )$-parameter plane. 

\begin{figure}[!t]
   \centering
   \includegraphics[width=.49\textwidth]{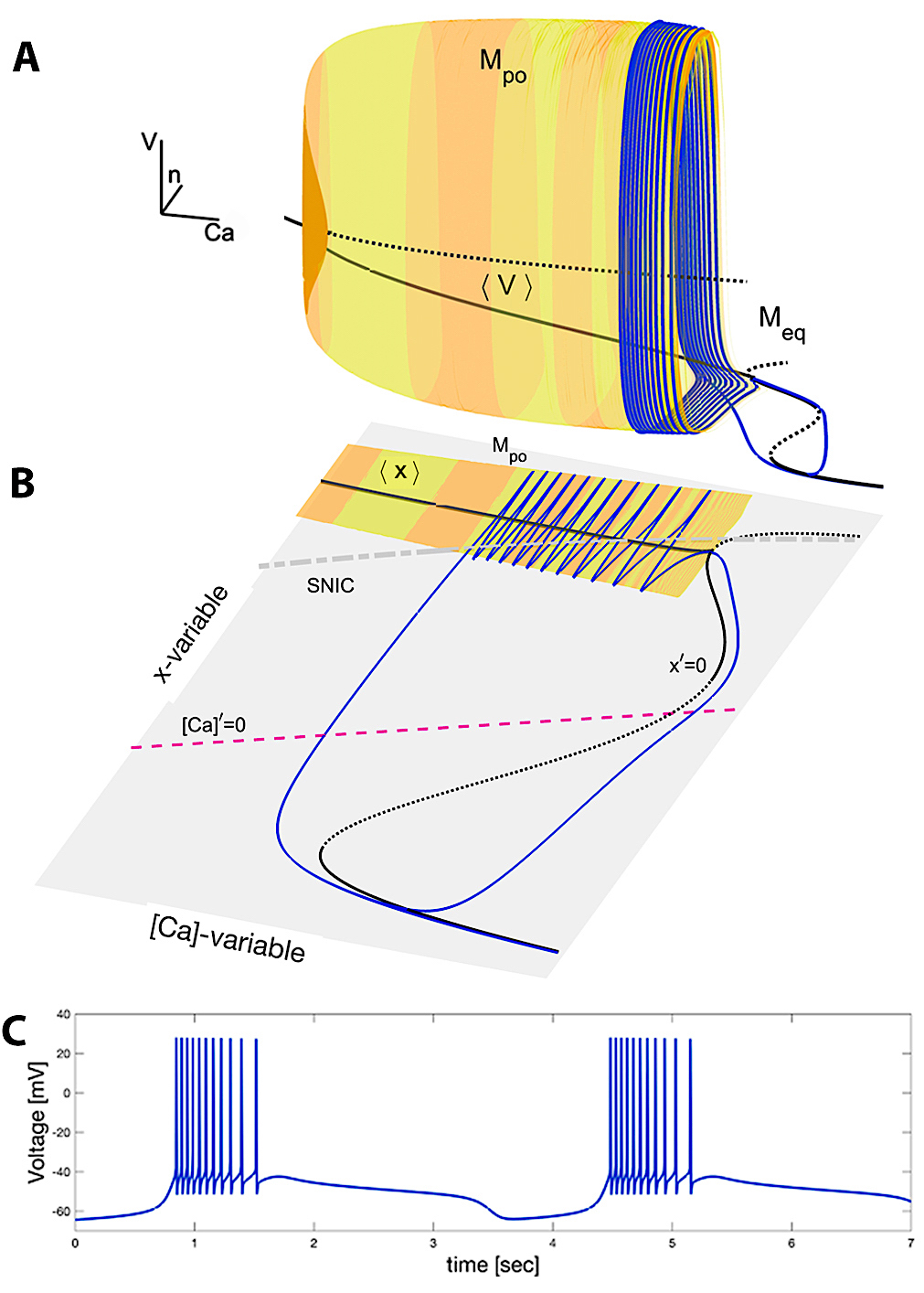}
   \caption{Panel A: 2D critical manifold $M_{\rm po}$ (of a cylindrical shape) is constituted of tonic-spiking periodic orbits in the $([\rm Ca],n,V)$-phase subspace of the Plant model at $\Delta_x=-1.7$mV and $\Delta_{\rm Ca}=15$mV. The space curve $\langle V \rangle$, representing  the time averages of fast oscillations, originates through a sub-critical AH-bifurcation point and terminates through the saddle-node point where it touches the top fold of a bent $M_{\rm eq}$-curve (with solid-stable and dotted-unstable sections) made of equilibria of the model. The blue trajectory, turning around the $M_{\rm po}$-surface and sliding onto and along the $M_{\rm eq}$-curve, and back to $M_{\rm po}$  is a bursting (periodic) orbit, whose voltage trace is shown in Panel~C. Panel B: Projections onto the slow $([\rm Ca],\,x)$-subspace of the Plant model reveal an intrinsic hysteresis within the slow dynamics, which is necessary for endogenous bursting due to two stable overlapping branches of the average nullcline $\langle x \rangle$ and the x-nullcline $x'=0$, which represent the $\langle V \rangle$- and $M_{\rm eq}$ manifolds, resp. Panel C: Voltage trace showing bursting activity with alternating episodes of hyperpolarized quiescent transients and trains of fast spikes.}
   \label{fig:Plant_burster}
\end{figure}

\begin{figure}[!]
   \centering
   \includegraphics[width=.4999\textwidth]{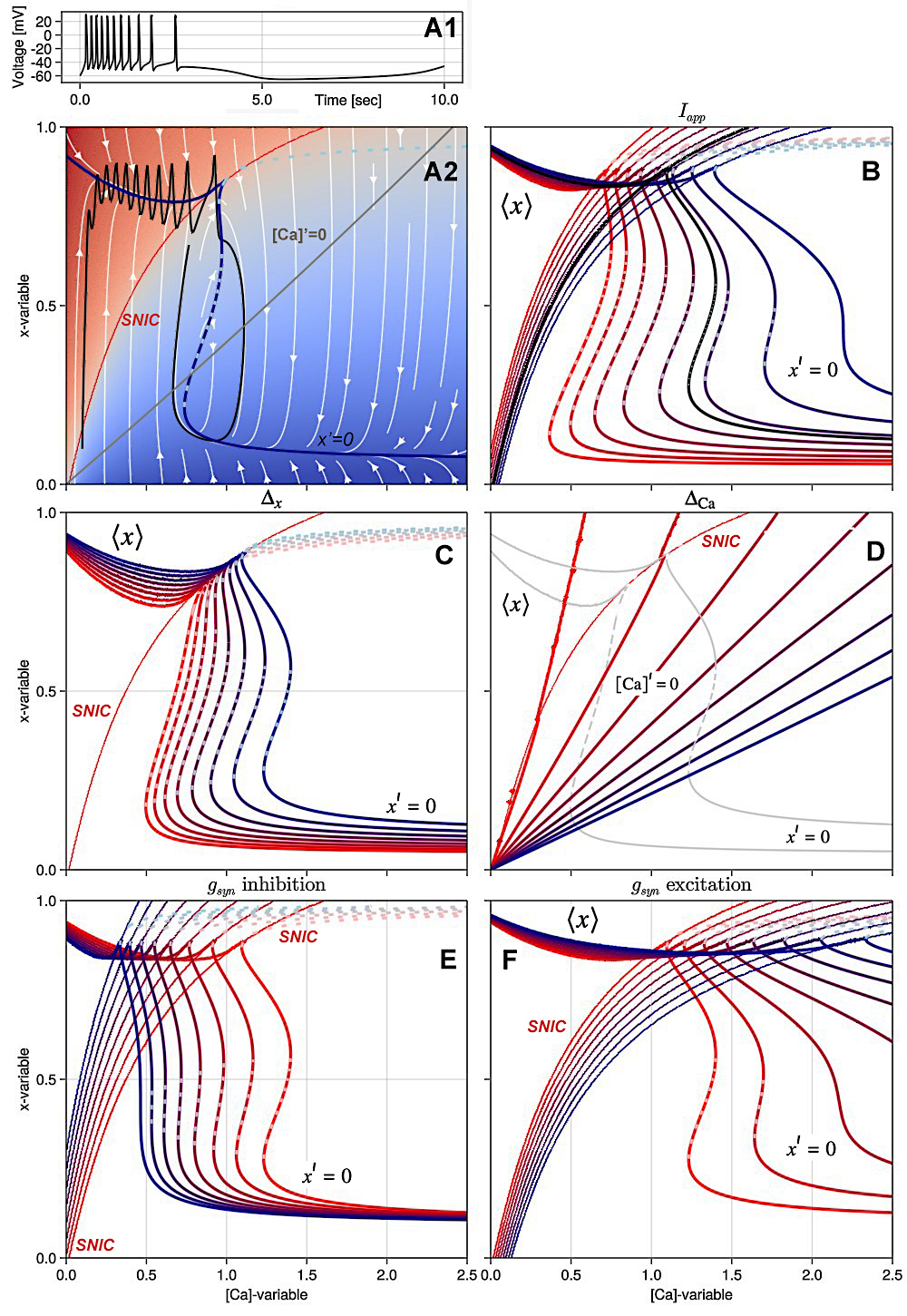}
   \caption{Reconstructions of the slow $([\rm Ca],\,x)$-phase plane as the parameters of the model are varied. Panel~A1: Voltage trace corresponding to Panel~A2. Panel~A2: Transient spiking trajectory in the $([\rm Ca],\,x)$-phase plane at $\Delta_{\rm x} = -4$mV and $\Delta_{\rm{Ca}} = -100$mV. The white stream-plot illustrates the flow/vector field on the plane. Indicated lines are: the average nullcline $\langle x \rangle$ (between the spiking orbit) (blue line); the stable/unstable branches of the equilibrium $x$-nullcline given by $x'=0$ are (dark blue line solid/dotted resp.); the nullcline $[\rm Ca]'=0$ (grey line), and the SNIC bifurcation curve (red line). The background heat-map in red and blue indicates the average high and low quantities of the membrane voltages resp. in $([\rm Ca],\,x)$-phase plane for fixed $[\rm Ca]$ and $x$-values. Panel~B: Evolutions of the SNIC-curve and the nullcline $x'=0$ featuring the pronounced hysteresis (overlap) as an external current $I_{ext}$ ranges from -0.1 (red lines) to 0.05 (blue lines), relative to the black curves corresponding to $I_{app} = 0$. Panel~C: Decreasing the $\Delta_x$-parameter from 0 to -4 shifts the nullcline $x'=0$ to the right, creates a stable section on it above the first knee-point, and illuminates it overlap the average nullcline $\langle x \rangle$. Panel D: Increasing the $\Delta_{\rm Ca}$-parameter from -100 through 350, turns the nullcline $[\rm Ca]'=0$ clockwise and shifts its intersection with the x-nullcline $x'=0$ across the SNIC-curve; the model transitions from tonic-spiking to quiescence at $\Delta_x=-4$mV (top nullcline $x'=0$), or when $\Delta_x=0$ from tonic-spiking to bursting (bottom nullcline $x'=0$). Panels~E and F demonstrate the rearrangements of the nullclines as the maximal conductance $g_{syn}$ of the inhibitory and excitatory currents are increased 0.001 and 0.01 resp. with $E_{rev}=-70$mV and $E_{rev}=30$mV resp.; solid/dotted lines represent stable/unstable branches of the equilibrium x-nullcline $x'=0$ in the $([\rm Ca],\,x)$-phase plane.}
   \label{fig:cascade}
\end{figure}

Let us outline what transformations or bifurcations occur on the transition routes throughout the borders  between the activity types. Note first that the transition route between tonic-spiking and bursting     activity, say as the $\Delta_{\rm Ca}$-parameter is increased (at some fixed $\Delta_x=0$) cannot be  described by a single bifurcation but a series of consecutive nonlocal bifurcations. These include a period-doubling cascade resulting in the onset of chaotic dynamics occurring within a narrow (grey) region in the parameter plane through which round periodic orbits for tonic-spiking activity morph into two time-scale bursting orbits composed of alternating episodes of fast spike trains and slow quiescent phases. This region is adjacent to a local saddle-saddle bifurcation curve (black line). Unlike a saddle-node bifurcation through which saddle and stable equilibrium states merge and vanish, or decouple, this bifurcation gives rise to two saddles emerging nearby.

\begin{figure*}[!t]
   \centering
   \includegraphics[width=.999\textwidth]{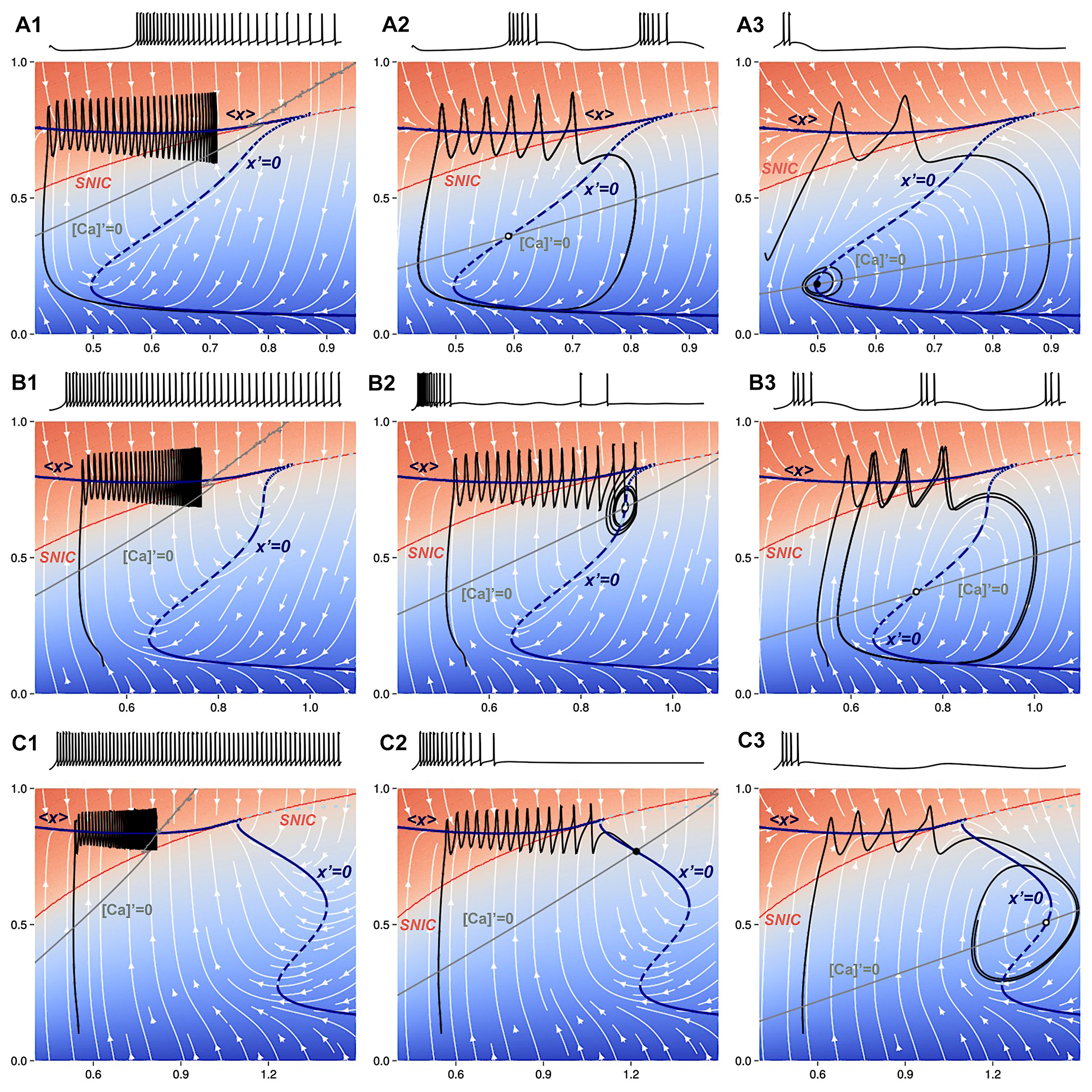}
   \caption{Various types of neural activity as depicted in the voltage traces corresponding in the slow $([\rm Ca],\,x)$-phase plane superimposed with the white stream-plot for selected $\Delta_x$- and $\Delta_{C\rm a}$-parameter values.  Shown are the following nullclines:  average $\langle x \rangle$ (blue line), $x'=0$ with stable/unstable branches (solid/dashed dark-blue lines), $[\rm Ca]'=0$ (grey line), along with the SNIC-curve (red line). 
   The top panels show an intermediate bursting regime at $\Delta_x = 0$. Panel~A1 shows tonic-spiking activity at $\Delta_{ \rm{Ca}} = -60$mV with a latent hysteresis due to overlap of the $x$-nullclines: average  $\langle x \rangle$ and the low branch of the equilibrium one $x'=0$. Panel~A2 shows the original Plant burster configuration at $\Delta_{ \rm{Ca}} = 0$mV. Panel~A3 shows a trajectory near the subcritical Andronov-Hopf bifurcation with decaying subthreshold oscillations near the lower knee point of the nullcline $x'=0$ at $\Delta_{ \rm{Ca}} = 120$mV.
   The middle panels show transient chaos en route to bursting at $\Delta_x = -1.4$mV. Panel~B1 shows tonic-spiking activity at $\Delta_{ \rm{Ca}} = -60$mV. Panel~B2 shows transient chaos for $\Delta_{\rm{Ca}} = -32$mV before a subcritical hopf bifurcation.  Panel~B3 shows a bursting trajectory at $\Delta_{ \rm{Ca}} = 40$mV after the unstable limit cycle disappears.
   The bottom panels show no occurrence of bursting for the $\Delta_x = -4$ where the neuron model transitions from tonic-spiking (Panel~C1) at $\Delta_{ \rm{Ca}} = -100$mV, directly  to the quiescent state (Panel~C2) at $\Delta_{ \rm{Ca}} = 0$mV,  and  finally to subthreshold oscillations (Panel~C3) at $\Delta_{ \rm{Ca}} = 105$mV with the corresponding limit cycle emerging  and collapsing sequentially through two supercritical Andronov-Hopf bifurcations near the high and low knee points respectively.}
   \label{fig:phase_examples}
\end{figure*}

For some fixed $\Delta_{\rm Ca}$, a (vertical) transition route from hyperpolarized quiescence to bursting in the bifurcation diagram, as the $\Delta_x$-parameter is increased, includes a local Andronov-Hopf (AH) bifurcation through which a stable equilibrium state representing the neural quiescence loses its stability to becomes unstable in the bursting region in the $\left ( \Delta_{\rm Ca}, \, \Delta_x \right )$-parameter plane. This AH-bifurcation can be sub- or super-critical on the opposite sides of the bifurcation curve divided by a codimension-2 point, known as a Bautin \footnote{N.N.~Bautin examined a structure of such a point where the criticality of this bifurcation alters from sub to super, and found that its unfolding includes an additional curve corresponding to a saddle-node bifurcation of periodic orbits~\cite{book} point (BP). Note that the occurrence of such a point in various neuronal models is a precursor of a possible transition to bursting through the torus bifurcation~\cite{ju2018bottom}.} point (BP). A bottom-up route to bursting on the right from the BP-point leads first to a gradual onset of stable sub-threshold oscillations of a small amplitude in voltage traces, which after increasing in size gain sequentially more spike to transform in fully developed bursting at larger $\Delta_x$-values. A similar route to the left from the BP-point is more complicated as it gives rise to complex bursting with chaotic sub-threshold oscillations or with unpredictably varying trains of spikes, or both. Moreover, bursting can also co-exist with a stable quiescent state within a small attraction basin, which manifests bistability in the neuron model. A few of these transitions to regions of interest are illustrated in Fig.~\ref{fig:phase_examples}. Our plan is to provide a detailed analysis of these transitions between neural activity types in a forthcoming paper.           
  
\subsection{Phase space dissection of the Plant burster} 

Slow-fast decomposition makes it feasible to visualize and interpret the dynamics in the full 6D phase space (with the $h$-current) of the Plant burster by projecting it onto the phase plane to analyze all pivotal bifurcations occurring in its slow 2D subsystem ~(\ref{SlowSub1}-\ref{SlowSub2}). To interpret the dynamics and its transformations we introduce the following four curves in the $([\rm Ca],x)$-phase plane: the $x$-nullcline, $ [\rm Ca]$-nullcline, the average $\langle x \rangle$ nullcline, and the bifurcation curve abbreviated as SNIC\footnote{SNIC (Saddle-Node-on-Invariant-Circle) stands for a homoclinic bifurcation of a saddle-node periodic orbit of codimension-one, which was first examined in a plane by Andronov and Leontovich \cite{andronov1937some,andronov1938, TB1, TB2}, and further generalized by L.P.~Shilnikov for ${\mathbf R}^n$-case  in his Ph.D.~\cite{Sh62, Sh63}, see also Refs.\cite{book,shilnikov2004some,Sciheritage,LPbook17} and references therein. The feature of a SNIC bifurcation is that it unfolds with the onset of either a stable periodic orbit of long period emerging from a homoclinic orbit to a saddle-node, or a stable equilibrium.} and explain how their interplay affects the overall dynamics of the given neuron model.

Figure~\ref{fig:Plant_burster} illustrates the geometry of the phase space of the original Plant burster. Its Panel~A represents a pseudo-projection \footnote{it is not a true projection, because the fast subsystem is restricted to its average, which is a cut through the full 6D phase space.} 
onto the 3D $([\rm Ca],n,V)$-subspace, which is  super-imposed with the $([\rm Ca],x)$-phase plane shown in Panel~B. The meaning of this bifurcation is that for variable values above the SNIC-curve in the $([\rm Ca],\,x)$-plane, the fast subsystem exhibits oscillatory, tonic-spiking activity, whereas below it becomes hyperpolarized quiescent. To find the SNIC-curve, one has to find and follow the saddle-node bifurcation in the fast subsystem~Eqs.~(\ref{eq:voltage})--(\ref{eq:hcurrent}) of the Plant model while varying the $x$- and $[\rm Ca]$-variables as two control parameters.

The projection in Fig.~\ref{fig:Plant_burster}A depicts a pair of the critical or slow-motion manifolds: a 2D cylinder-shaped surface $M_{\rm po}$ (painted with alternating yellow/orange stripes) of periodic orbits originating from and terminating on a 1D multi-folded (black) space curve $M_{\rm eq}$ of equilibrium states. In addition, both manifolds are overlaid with a (blue) bursting periodic orbit, the voltage trace of which is presented in Fig.~\ref{fig:Plant_burster}C. Such 2D and 1D critical manifolds can be found, for example, by following the branches of periodic orbits and equilibrium states in the fast subsystem of the Plant model using numerical continuation by freezing slow variables and treating them as control parameters in the slow-fast dissection. 

Figure~\ref{fig:Plant_burster}B also depicts these critical manifolds projected on $([\rm Ca],x)$-phase plane. Here, the $x$-nullcline given by the condition $x'=0$ corresponds to the 1D manifold $M_{\rm eq}$ in the full phase space.  Figure~\ref{fig:Plant_burster}B also shows the SNIC-curve.
 
The (black) solid space curve labeled by $\langle V \rangle$ in Fig.~\ref{fig:Plant_burster}A and its equivalent denoted by $\langle x \rangle$ in Panel~B  represent the ``gravity center'' of the critical manifold $M_{\rm po}$. This manifold $M_{\rm po}$ is foliated by the periodic orbits representing tonic-spiking activity in the neuron model. By averaging the fast coordinates, $V_{\rm po}(t)$, $ x_{\rm po}(t)$, etc., of every periodic orbit over its period: 
\begin{equation}\label{aver}
\langle V \rangle= \frac{1}{T} \int_0^T V_{\rm po}(t) dt  \quad \mbox{and}   \quad  \langle x \rangle = \frac{1}{T} \int_0^T x_{\rm po}(t) dt, 
\end{equation}  
one can find the points that constitute such curves of averages in the phase space. This curve begins from the point where $M_{\rm po}$ collapses into the $M_{\rm eq}$-manifold and terminates at the top fold on the given 1D manifold, throughout which the bent SNIC-curve passes. The reader will find more details on the averaging approach, including the notion of average nullclines to locate a periodic orbit in the phase space and corresponding to tonic-spiking activity in the neuron model in the Appendix~II\hyperref[sec:appendix2b]{B} below. 
  
We now focus on the $([\rm Ca],\,x)$-phase plane and specifically on the interplay of the following two nullclines: the $x$-nullcline where $x'=0$, and the [$\rm Ca$]-nullcline where $[\rm Ca]'=0$. These nullclines determine the vector field in the phase plane that directs solutions of the slow subsystem as depicted in Fig.~\ref{fig:cascade}A2. By the meaning of a nullcline, the vector field decreases/increases in the $x$-direction above/below the $x$-nullcline; its stable and unstable sections are indicated by solid and dotted intervals, where $x'<0$ or $x'>0$, respectively, on it. 
 
The $x$-nullcline in Fig.~\ref{fig:cascade}D is overlaid with the (red) line representing the slowest $[\rm Ca]$-nullcline such that below/above it the [$\rm Ca$]-variable decreases/increases, respectively. One can observe from  Eq.~(\ref{SlowSub2}) that the slope and the position of the $[\rm Ca]$-nullcline is determined by the $\Delta_{\rm Ca}$-parameter; namely, decreasing $\Delta_{\rm Ca}$ makes the $[\rm Ca]$-nullcline shift leftwards in the $([\rm Ca],\,x)$-plane and vice versa. 
  
Let us next carry out a geometric analysis of these nullclines and how they determine the slow dynamics of the neuron model. By construction, a point, where both nullclines cross in $([\rm Ca],\,x)$-plane, and where $x'=0$ and $[\rm Ca]'=0$ is an equilibrium state of the slow subsystem. It can be stable or unstable/repelling, depending on whether the $[\rm Ca]$-nullcline intersects a stable (solid) or unstable (dotted) branch of the $x$-nullcline. A tangency of the nearly straight $[\rm Ca]$-nullcline with the bending $\Sigma$-shaped $x$-nullcline corresponds to a saddle-node or a saddle-saddle bifurcation of equilibrium states in the model in the $(\Delta_{\rm Ca},\,\Delta_{x})$-parameter plane, see the bifurcation diagram in Fig.~\ref{fig:bf_diagram}. This occurs when the $[\rm Ca]$-nullcline, turning clockwise as the $\Delta_{\rm Ca}$-parameter increases, becomes tangent to the unstable (top) branch of the $x$-nullcline (as demonstrated in Fig.~\ref{fig:cascade}D). Recall that the transverse crossing of the $[\rm Ca]$-nullcline through a fold of the $x$-nullcline, where its stable and unstable branches merge, correspond to an Andronov-Hopf bifurcation that occurs in the slow subsystem of the neuron model. Here we refer the reader back to the bifurcation diagram in Fig.~\ref{fig:bf_diagram}.       
  
The geometric configuration of the slow nullclines in the $([\rm Ca],\,x)$-phase plane determines dynamics of the slow subsystem and, consequently, the behavior of the full model. To be attracting, a periodic orbit or an equilibrium state must be stable in both slow and fast subspaces of the phase space of the Plant burster. Whenever either is repelling (or unstable) in the slow $([\rm Ca],\,x)$-plane, then it will become of the saddle type, and thus barely observable in the full system, see Refs.~\cite{shil2004homoclinic,Mmo2005, Shilnikov2008a, Shilnikov2012, ju2018bottom}.
  
Both the slow motion manifolds $M_{\rm eq}$ and $M_{\rm po}$, and their constituent attractors in the fast $(V,n,h)$-subspace are stable in any HH-type or phenomenological models as they produce observable spike generation and resting states within a proper parameter range. According to this classification, there are three possibilities: i) a stable equilibrium state located below the SNIC-curve in $([\rm Ca],\,x)$-plane is a steady state in the whole system and corresponds to hyperpolarized quiescence in the neuron model ii) a stable equilibrium located above the SNIC-curve in the $([\rm Ca],\,x)$-plane is a stable periodic orbit in phase space that corresponds to tonic-spiking activity iii) an unstable equilibrium state located below the SNIC-curve corresponds to a saddle point, which has a pair of complex conjugate eigenvalues with positive real parts. This last case can correspond to chaotic dynamics, periodic bursting, or subthreshold oscillations, the discussion of which is beyond the scope of this paper.
  
It is evident that the geometric configuration based on the average $\langle x \rangle$ nullcline connecting to the equilibrium nullcline $x'=0$ in the $([\rm Ca],\,x)$-phase plane is reminiscent of the single cubic nullcline found in a relaxation oscillator and its biological interpretation -- the Fitzhugh-Nagumo neuron.
The hysteresis in this construction, marked by two overlapping branches, is a necessary condition for oscillations in the Fitzhugh-Nagumo neuron, which would correspond to endogenously bursting oscillations in the Plant model. However, this is not sufficient yet as both overlapping branches are to be transient for oscillatory solutions and hence should not have stable equilibria on them. Figure~\ref{fig:Plant_burster}B depicts such a configuration where the nullcline $[\rm Ca]'=0$ crosses the nullcline $x'=0$ throughout its unstable branch. This results in the onset of a stable limit cycle (blue line) in the $([\rm Ca],\,x)$-phase plane, which corresponds to a bursting periodic orbit and shown in Panel~A where the orbit turns around the tonic-spiking manifold $M_{\rm po}$, slides onto the quiescent manifold $M_{\rm eq}$, and back to $M_{\rm po}$ and so forth to generate the voltage trace in Panel~C of the same figure. 

\subsection{Transitions due to intrinsic and external parameters}  

In addition to the intrinsic parameters, $\Delta_x$ and $\Delta_{\rm Ca}$ governing the cellular
dynamics, let us introduce external parameters to perturb isolated neurons. These are a constant applied current, $I_{app}$, and a synaptic drive or current given by $I_{syn} = g_{syn}\,(E_{rev}-V)$, which can be inhibitory when $V(t) > E_{rev}$, where $E_{rev}=-80$mV or less, or excitatory with $E_{rev}=+40$mV.  

The panels of Fig.~\ref{fig:cascade} demonstrate  the overall effects of individual variations of  four control parameters $I_{app}$, $\Delta_x$, $\Delta_{\rm Ca}$, and $g_{syn}$ on the geometric organization and its rearrangements of the slow $([\rm Ca],\,x)$-phase plane.

For illustration we begin with Fig.~\ref{fig:cascade}A2 as a reference to introduce the flow lines (shown as white lines with arrows) of the vector field to direct a trajectory (black line) in the phase plane, as well as to discuss a novel computation approach to derive the vector field,  and to approximate the locations of the SNIC-curve and the average nullcline $\langle x \rangle$ (the  blue line between the fast spikes in the projection) alternatively in the phase space of the neuron model.   

Let us start with the effect caused by $\Delta_x$-variations. First, recall from the bifurcation diagram in Fig.~\ref{fig:bf_diagram} that the Plant model morphs from  the endogenous burster to a no-burster as $\Delta_x$ is decreased, Therefore the $([\rm Ca],\,x)$-phase plane must be re-arranged to break down or lose the initial hysteresis due to the overlapping of the nullclines for periodic orbits and equilibria. This is {\em de-facto} supported by Fig.~\ref{fig:cascade}C that decreasing $\Delta_x$ makes  the nullcline  $x^\prime = 0$ bend leading to the emergence of an upper stable branch right below the SNIC-curve. In addition, one can see the $\Sigma$-shaped nullcline $x'=0$ has shifted to the right, thus eliminating an overlap with $\langle x \rangle$. This is indicative that at $\Delta=-4$mV, the neuron model can only transition from the tonic-spiking activity to the hyperpolarized quiescent state as $\Delta_{\rm Ca}$ is varied.  

The effect of $\Delta_{\rm Ca}$-variations is depicted in Fig.~\ref{fig:cascade}D and shows that $\Delta_{\rm Ca}$ effectively changes the slope of the $[\rm Ca]$-nullcline. One can see from the panel that changing the slope of  the $[\rm Ca]$-nullcline shifts its intersection point with the nullcline $x^\prime = 0$. The panels shows two different positions and shapes of the $x$-nullclines for different values: $\Delta_x=0$mV and  $\Delta_x=-4$mV. We will come back to discuss the ongoing bifurcations caused by the shape transformations and $\Delta_{\rm Ca}$-variations below. 

Figure~\ref{fig:cascade}B shows the two effects on the rearrangements of the $x$-nullcline and the SNIC-curve by $I_{app}$-variations. For reference, the original SNIC-curve and the nullcline $x'=0$ are shown as bold black curves. One can see from the panel that applying a positive $I_{app}$-current further bends and translates the SNIC-curve to the right, thereby extending the tonic-spiking domain and increasing the overall excitability of the neuron. Note that with the nullcline $x^\prime = 0$ being shifted rightwards, the overlap needed for the hysteresis is gone as well. Application of negative $I_{app}$-currents produces the opposite effect by decreasing excitability of the neuron and shrinking the oscillatory domain. Meanwhile the nullcline $x' = 0$ is shifted leftward and further bent with lower stable hyperpolarized branch.  

Figures~\ref{fig:cascade}E and F demonstrate the effects of the synaptic drive $I_{syn} = g_{syn}\,(E_{rev}-V)$. While its local effect on the SNIC-curve is qualitatively similar to the application of the constant current, the overall outcome of the synaptic drive on slow dynamics is profoundly different and significant. One can see from Panel~E that increasing the inhibitory current (with $E_{rev}=-70$mV) shifts the nullcline $x^\prime = 0$ to the left, as well as straighten its shape and eliminates its upper knee points.  Figure~\ref{fig:cascade}F illustrates how the application of an increasing excitatory current translates the location of the SNIC-curve and reshapes the nullcline $x'=0$, eliminating hysteresis and hence bursting activity.

There is a simple explanation for the different effects of $I_{app}$ and $I_{syn}$ currents. The latter term can be factored into two sub-terms: constant $g_{syn}\,E_{rev}$ which can be treated as some constant current $I_{app}$, positive or negative, depending in the sign of $E_{rev}$ and a time-variable term $-g_{syn}\,V(t)$, which also becomes fixed at a voltage steady state, but its value varies along the steady state nullcline $x^\prime = 0$. As we have seen above that this term always makes the hysteresis shrink by a shear-like transformation of the nullcline, thus overwhelming the opposite effects induced by $I_{app}$. We stress that the difference between applications of the constant current $I_{app}$ and and synaptic current $I_{syn}$ has profound implications for the interpretation of electrophysiological experiments. However, for this paper, we only offer a word of caution, namely applying constant currents can change the underlying slow dynamics of cells which can also create the possibility of measuring artifacts that are not anyhow relevant to the normal functioning of a neural circuit or intrinsic cell behaviors.

 
To understand the difference in the parameter regimes of the adopted Plant model (see its bifurcation diagram in Fig.~\ref{fig:bf_diagram}), choose $\Delta_{\rm Ca}$ as a bifurcation parameter and explore how its variations are correlated with specific bifurcations and how they underlie the transitions between neural activity types.  

One can observe from the panels in Fig.~\ref{fig:phase_examples} that depending on $\Delta_x$-values the equilibrium state nullcline $x^\prime = 0$ may have up to three different folds on it. Each fold is a knee point in the $[\rm Ca]$-direction and corresponds to a change in stability of the adjacent sections, stable (sold line) and unstable (dotted line) on the nullcline $x' = 0$. It is a well-known fact that such a knee point corresponds to a saddle-node (or fold) bifurcation through which two equilibria, stable and repelling, merge and vanish in a slow subsystem where $[\rm Ca]$-variable becomes frozen to be treated as a control parameter. We note here that these two equilibria can be both saddles in the full model as well.    

We consider initially the case where the neuron model exhibits tonic-spiking activity as shown in Fig.~\ref{fig:phase_examples}A1. This activity is due to a single stable periodic orbit, to which a phase trajectory (shown as a black line) in its attraction basin converges, see the $([\rm Ca],\,x)$-phase plane in Fig.~\ref{fig:phase_examples}A1. Note from this panel that after the trajectory crosses above the SNIC-curve, it starts zigzagging (spiking) and shifting toward the stable orbit next to the nullcline $[\rm Ca]'=0$ (grey line).

As $\Delta_{\rm Ca}$ increases further, the nullcline $[\rm Ca]'=0$ turns clockwise past the SNIC-curve. To proceed with the analysis, we must consider two cases separately. 
\begin{itemize}
\item Case 1: the average nullcline $\langle x \rangle$  connects to an unstable  branch of the nullcline $x'=0$. Then, tonic-spiking activity transforms into bursting activity as occurs in the original Plant model with $\Delta_x = 0$mV. This case is illustrated in Figs.~\ref{fig:phase_examples}A2. 
\item Case 2:  the average nullcline $\langle x \rangle$  connects to a stable branch of the nullcline $x'=0$. Then tonic-spiking activity transitions to the quiescent state directly as occurs at the level $\Delta_x = -4$mV. This case is illustrated in Figs.~\ref{fig:phase_examples}B2. 
\end{itemize}

Figures~\ref{fig:phase_examples}A1-A3 demonstrate three snapshots of the $([\rm Ca],\,x)$-phase plane transformations beginning with tonic-spiking activity as $\Delta_{\rm Ca}$ is increased at $\Delta_x =0$mV. Recall that in terms of the fast subsystem the stable periodic orbit vanishes through the SNIC bifurcation that gives rise to a stable equilibrium state. In terms of the slow subsystem this transition occurs where the nullcline $[\rm Ca]'=0$  in the $([\rm Ca],\,x)$-phase plane crosses the cusp where the corresponding stable average nullcline $\langle x \rangle $ merges with an unstable nullcline $x' = 0$. Soon after the nullcline $[\rm Ca]'=0$ is decreased further to transversally cross the middle, unstable (dotted, highlighted) section of the nullcline $x'=0$, the model starts bursting ( as seen in \ref{fig:phase_examples}A2) due to periodic alternations between tonic-spiking episodes characterized by increasing $[\rm Ca]$-concentration, and quiescence phases during which $[\rm Ca]$ decreases. With a further decrease of $\Delta_{\rm Ca}$ the nullcline intersection lowers below the bottom knee point to transversally cross the stable lower branch of the nullcline $x' = 0$, where the neuron becomes hyperpolarized quiescent. This transition is associated with two stages: 1) the ``bursting'' limit cycle initially shrinks in size below the SNIC-curve so that the neuron starts exhibiting subthreshold oscillations of small amplitude. Next, 2) the stable limit cycle collapses at the equilibrium state near the knee through a supercritical AH-bifurcation.

Figures~\ref{fig:phase_examples}B1-B3 represent three parallel transition stages as $\Delta_{\rm Ca}$ is increased at $\Delta_x = -1.4$mV, before the Bautin point. The system starts off as a tonic spiker as before Fig.~\ref{fig:phase_examples}B1. As the intersection between the two slow nullclines reaches the SNIC-curve, the system undergoes a subcritical AH bifurcation creating a stable equilibrium bounded by an unstable limit cycle (Fig.~\ref{fig:phase_examples}B2). On the exterior of the limit cycle, there is transient bursting. As $\Delta_{\rm Ca}$ increases further, the unstable limit cycle and the basin of attraction of the stable equilibrium grow and shrink, until the unstable manifold of the unstable orbit separates from the chaotic bursting, leading to bistability. As the unstable limit cycle finally disappears in a second a subcritical AH bifurcation, stable bursting emerges Fig.~\ref{fig:phase_examples}. Increasing $[\rm Ca]'=0$ further sends the intersection below the bottom knee point leading to quiescence ( Fig.~\ref{fig:phase_examples}A3). 

Figures~\ref{fig:phase_examples}C1-C3 demostrate the changing behavior as $\Delta_{\rm Ca}$ is increased at $\Delta_x = -4$mV. As the intersection between the two slow nullclines lower below the SNIC-curve to cross the stable (solid) section of the nullcline $x' = 0$, the neuron becomes quiescent (Fig.~\ref{fig:phase_examples}B2). As $\Delta_{\rm Ca}$ increases further, a supercritical AH-bifurcation  marks the onset of subthreshold oscillations as the corresponding limit cycle does not reach the SNIC-curve in the $([\rm Ca],\,x)$-phase plane(Fig.~\ref{fig:phase_examples}B3). Finally, as the $[\rm Ca]'=0$ lowers bellow the bottom knee point, the subthreshold oscillations are seized by the hyperpolarized quiescent state, as Fig.~\ref{fig:phase_examples}A3 demonstrates. 

To conclude this section, we recap how can apply our new knowledge of the nullclines, and how they are affected by the control parameter variations to transform the bursting dynamics of the model (Fig.~\ref{fig:Plant_burster}). First, evaluate a range of the control parameters. Second, describe a bifurcation sequence and underlying transitions in the neuron model to exhibit periodic bursting. 

The key is in the shape of the $x$-nullcline. It is easy to see that decreasing $\Delta_{\rm Ca}$-parameter makes the nullcline $[\rm Ca]'=0$ turn counter-clockwise through some angle until it transversally crosses the stable (solid line) section of the nullcline $x'=0$ (Fig.~\ref{fig:Plant_burster}B). The intersection point is therefore a stable equilibrium state of the model. Note that the given stable section is directly connected to the $M_{\rm po}$-manifold. So, further decreasing $\Delta_{\rm Ca}$-parameter makes the nullcline $[\rm Ca]'=0$ turn counter-clockwise more until it transversally crosses the average $\langle x \rangle$ nullcline in the $([\rm Ca],\,x)$-phase plane in Fig.~\ref{fig:Plant_burster}B. By fitting this bifurcation sequence with the diagram in Fig.~\ref{fig:bf_diagram} one can deduce that the transition route began in the bursting (white) domain, through the quiescent (green) domain and finished in the tonic-spiking (yellow) domain. With saying that a qualitative evaluation of the $\Delta_x$-parameter is around $-2$mV while $\Delta_{\rm Ca}$ was decreased from $+20$ through $-60$mV.  

\begin{itemize}
\item
The bifurcation diagram in Fig.~\ref{fig:bf_diagram} suggests that there are three different transition routes through which the neuron model can undergo as $\Delta_{Ca}$ is increased within [-80,\,50]mV range: transition i) from tonic-spiking to bursting activity directly at level $\Delta_x=0$, or ii) additionally throughout the quiescent state at $\Delta_x=-2$, or iii) from tonic-spiking to the quiescence only at $\Delta_x=-4$.
\end{itemize} 
 
Another important observation is that the Plant model demonstrates bursting activity only if the tonic-spiking manifold remains transitive for the solutions that quickly turn around it (to spike) while slowly drifting towards its right end at the higher $[\rm Ca]$-values. As such, the burst duration, and the number of spikes per burst, as well as the interspike intervals and the interburst duration are also directly determined by the slow calcium dynamics, see Eq.~(\ref{SlowSub1})-(\ref{SlowSub2}). Meanwhile, it becomes evident that the change rate $[\rm Ca]'$ can implicitly be regulated by the position of the calcium nullcline relative to that of the curve $\langle x \rangle$ for spikes and the nullcline $x'=0$ for the quiescent phase, see Fig.~\ref{fig:Plant_burster}. This raises the question concerning how external perturbations, such as synaptic inhibition and excitation, can affect the characteristics of evoked bursting and emerging network hysteresis, namely what their range, variation and robustness can be, for example.   
  
\subsection{Neuron bursting in response to synaptic inhibition}   

One of the key features of the swim CPGs in the sea slugs, {\it Melibe leonina} and {\it Dendronotus iris}, is that their interneurons neither burst endogenously nor when perturbed by application of external currents. Experimentally they will remain quiescent, sporadically emit action potentials, or to generate tonic-spiking activity during swim episodes having received excitatory drive from sensory neurons. 
	
Here is a short list of the prerequisites derived from the experimezzntal observations concerning biological swim CPG interneurons that the mathematical model should meet and exhibit:

\begin{enumerate}
\item  no latent bursting;
\item a post-inhibitory rebound after hyperpolarization; 
\item slow spike frequency adaptation.    
\end{enumerate}
	
As we have numerically determined above that the adopted neuron model at $\Delta_x=-4$mV can only exhibit tonic-spiking and quiescent activity depending on the $\Delta_{\rm Ca}$-parameter value. In what follows we need to validate whether the adopted model meets the other two conditions as well.    

\begin{figure*}[!ht]
\centering
\includegraphics[width=0.85\textwidth]{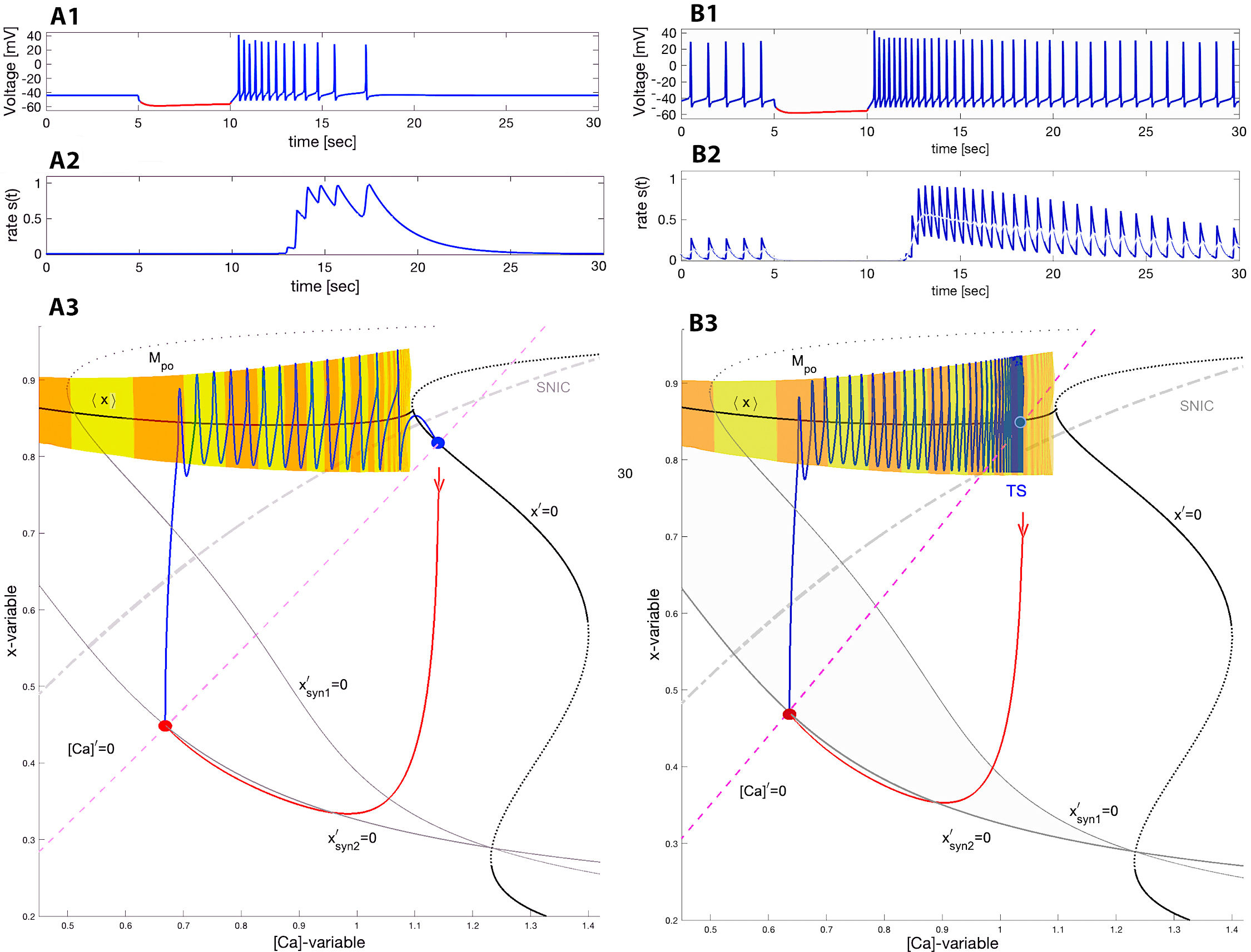}
\caption{Voltage trace showing quick post-inhibitory rebounds followed by the spike frequency adaptation in the neuron model returning to the stable quiescent state at ($\Delta_{\rm Ca}=-20$mV, $\Delta_{x}=-4$mV) in Panel A1 or to tonic-spiking activity at ($\Delta_{\rm Ca}=-35$mV, $\Delta_{x}=-4$mV) (Panel B1) after it becomes hyperpolarized by an inhibitory {\em synaptic} current perturbation  $-g_{\rm syn} (V(t)+80)$. Panels~A2 and B2 showing the time evolution of the neurotransmitter release rates matching the spike frequency changes resulting in stronger summation in the logistic synapses (introduced in next section  Eq.~\ref{sigmoid}). Panels: A3 and B3 showing the $([\rm Ca],x)$-phase projection showing the tonic-spiking manifold $M_{\rm po}$, and the unperturbed $\Sigma$-shaped nullcline $x'=0$ superimposed with two perturbed ones $x'_{\rm syn1,2}=0$ straightened by the application of inhibitory {\em synaptic} current perturbations with $g_{\rm syn}=0.5$ and $1$ (left black line). (A3) Forced transition from the depolarized quiescent state  (blue dot) at the intersection point of the nullcline $[\rm Ca]'=0$ (pink dashed line) with a stable branch on $x'=0$ towards the stable quiescent state of the neuron in normal conditions, while the red dot on the nullcline $x'_{\rm syn2}$ at $g_{\rm syn}=1$  corresponds to a hyperpolarized steady state emerging temporarily due to the force by the inhibitory synaptic current pulse. Panel~B3 depicts the response to the same perturbation of the neuron model  transitioning back to the stable periodic orbit at $\Delta_{\rm Ca}=-35$mV, located on the manifold  $M_{\rm po}$ near the intersection point of the average curve $\langle x \rangle$ and the calcium nullcline $[\rm Ca]'=0$ above the SNIC-curve in the $([\rm Ca],x)$-projection.}
\label{fig:two_panels}
\end{figure*} 

Figures~\ref{fig:two_panels} A and B is demonstrate how the geometric properties of the model fulfill all three conditions. It illustrates a response of the quiescent (panel A with $\Delta_{\rm Ca}\ge -20$mV) and tonic-spiking (panel B with $\Delta_{\rm Ca} \le -35$mV) interneurons after they are temporarily hyperpolarized by a 5[sec]-long inhibitory synaptic current (episodes in red). The voltage traces in Fig.~\ref{fig:two_panels}A1 and B1 show that after the perturbation the neuron model exhibits a {\em fast} post-inhibitory rebound with a high spike frequency initially, followed by a {\em slow} adaptation while returning to their respective attractors. Note that the slow spike frequency adaptation observed in the voltage traces is driven by the slow $[\rm Ca]$-dynamics in the model. This is another feature of this neuron  model: the spike frequency is maximized at low $[\rm Ca]$. Moreover, we can calibrate the neuron model to fit its range of firing rates between 6-16Hz, as observed in experimental recordings.
	
To better understand how the neuron model meets conditions 2 and 3 above, consider the neuron model under perturbation caused by a pulse of the inhibitory synaptic current $I_{syn}=g_{\rm syn} (V(t)+80)$. The pulse duration of 5 seconds is long enough for the neuron model to converge onto a newly perturbed stable state, as can be seen from the voltage traces. In addition to the original, unperturbed nullcline $x'=0$, figures~\ref{fig:two_panels}A3 and B3 depict two additional perturbed nullclines, labelled $x_{\rm syn1}'=0$ and $x_{\rm syn2}'=0$, corresponding to two different $g_{\rm syn}$-values, 0.5 and 1 respectively. The intersection point (red dot) of the nullcline $[\rm Ca]'=0$ with the stable nullcline $x_{\rm syn2}'=0$ at $g_{\rm syn}=1$ is a stable equilibrium state of the inhibited neuron. Shown in red is a phase trajectory forced to quickly transition from the unperturbed stable equilibrium state in figure~\ref{fig:two_panels}A3 or periodic orbit figure~\ref{fig:two_panels}B3 towards the perturbed steady state (red dot). This steady state persists as long as the inhibitory current lasts.
 
As soon as the inhibition is removed, the neuron model responds with a fast post-inhibitory rebound (PIR) associated with the trajectory (blue line) in the $([\rm Ca],\,x)$-phase plane that takes off nearly vertically from the perturbed steady state towards the tonic-spiking manifold $M_{\rm po}$. Having landed onto $M_{\rm po}$, it slowly transitions back to its original state along the manifold with a gradually decreasing spike rate. Note that the decreasing size of the steps between spikes on the manifold $M_{\rm po}$ in the phase plane is deceptive, because orbits to the left of the $M_{\rm po}$ populate the phase plane less densely, even though they are slower. This is simply a consequence of exponential convergence to the attractor in the slow subsystem, and is not indicative of the speed of the periodic orbit in the fast subsystem. Figure~\ref{fig:two_panels} includes two middle panels B2 and A2: both display what the time-varying proportion $S(t)$ of synaptic channels that are open due to neurotransmitter release.

To summarize the discussion on the cellular properties:

\begin{itemize}
\item Quiescent and tonic-spiking neurons in isolation can produce an episodic burst after recovery from forced inhibition. 
\item The stronger the inhibition is, the greater spike frequency becomes in the post-inhibitory rebound due to smaller $[\rm Ca]$-values associated with the forced state in the neuron, and the more pronounced the slow spike frequency adaptation is in the corresponding voltage trace.
\item The duration of the post-inhibitory rebound and the adaptation speed are determined by the change rate $[\rm  Ca]'$ of the calcium concentration, as well as the distance to travel back to the initial state of the neuron, which is determined by the position of the nullcline $[\rm Ca]'=0$ (due to $\Delta_{\rm Ca}$) in the $([\rm Ca],\,x$)-phase plane.  
\end{itemize} 

In the following sections we will show how the cellular dynamics enhanced with a strong post-inhibitory rebound (PIR) and the slow spike frequency adaptation can coordinate with slow synaptic dynamics to generate emergent network-level bursting in two generic types of building blocks for rhythm-generating circuits.

\section{Modeling synaptic dynamics: from fast to slow} \label{sec:3}

In this section we introduce and discuss several representative models describing chemical synapses, both fast and slow, through which neurons are coupled to produce network-level oscillations, orchestrate phase relationship between them and maintain stability of emerging bursting rhythms. Slow synapses are of particular interest, as they occur on the same timescale as network oscillations and have been reported in the swim CPGs of the sea slugs. We introduce a new {\em logistic} model that can describe both delayed and slow synaptic summation, which can act as a high-pass filter. The last section of the paper demonstrates how neural models coupled with such logistic synapses can produce robust self-sustained bursting at the network level.

Synaptic models strongly resemble the cellular currents in the Hodgkin Huxley formalism. Analogously, the synaptic current $I_{syn}$ in the post-synaptic neuron is modeled as follows:  
\begin{equation}
\quad I_{syn} = g_{syn}\, S(t) \, (V_{post}(t)-E_{rev}), \label{Eqsyn}
\end{equation}
where $g_{syn}$ is a maximal conductance, $S(t) $ is a synaptic probability, $V_{post}$ is the membrane voltage in the post-synaptic neurons, and $E_{rev}$ is a synaptic reversal potential, which can be set at +40mV or -80mV for excitatory and inhibitory synapses, respectively. The diversity of approaches to synaptic modeling stems from the variety of ways in which the proportion variable $S(t)$, sometimes referred to as a normalized coupling function, is related to the value of presynaptic voltage. The choice of the model for the time evolution of $S(t)$ is particularly important for slow synapses, as the dynamics may be critical to network function including rhythmogenesis. In the neuroscience context, the synaptic proportion $S(t)$ corresponds to proportion of open synaptic channels in the postsynaptic neuron at the given instant. Below we overview several options for modeling $S$-dynamics for both fast and slow synapses. 
We begin by introducing the sigmoid  function $f_{\infty}(V)$, which is used in multiple models of neurons and synapses
\begin{equation}
f_{\infty}(V) = \frac{1}{1+e^{-k(V_{pre}-\Theta_{syn})}},  \label{finf}
\end{equation}
where the constant $k$ determines its derivative at $f_{\infty}(V) = 0.5$, the inflection point where $V=\Theta_{syn}$. Here $\Theta_{syn}$ is treated as the synaptic threshold typically set around $+20$mV in the middle of spikes, between the spike threshold around $-45$mV (sodium channels opens) and the spike peak $+40$mV. Values for $k$ are typically set somewhere in $k \in \left[0.5, 50\right]$, and in the limit $k \rightarrow \infty$ the function becomes a continuous adaptation of the Heaviside step function, where the function $f_\infty$ remains close to zero when $V_{pre}(t) < \Theta_{syn}$, and quickly jumps to 1 after the voltage in the presynaptic neurons exceed the synaptic threshold. After an action potential, the voltage $V_{pre}$ quickly drops down below the synaptic threshold, which in turn halts the neurotransmitter release with the rate close to zero. It is worth repeating that these ideas of instantaneous activation/inactivation for such $f_{\infty}$ functions were initially proposed in the HH-formulation. All the following synapse types we review below employ the notion of $f_{\infty}$ functions to describe time-varying probabilities in their models. 

\subsection{Synapse types}
\paragraph{Fast threshold modulation} (FTM). 
This simple paradigm~\cite{Wang-Rinzel-92,FTM} adequately describes fast synapses. In a FTM framework, the synaptic activity is turned ``on'' or ``off'' instantaneously. This directly utilizes the aforementioned $f_{\infty}$ function with the presynaptic membrane potential $V_{pre}$: 
\begin{equation}
S(t) = f_{\infty}(V_{pre}(t)). \label{ftm}
\end{equation}
 Such fast synapses happened to be useful for understanding synchronized neuronal activity, including bursting with inhibitory synapses, and multistability in small, weekly coupled neural networks~\cite{prl08, Shilnikov2008b, Jalil-Belykh-Shilnikov-10, Wojcik2011a,  Wojcik2014, schwabedal2016qualitative,kelley20202,collens2020dynamics, pusuluri2020computational}. Moreover, this assumption has the benefit of simplifying the network dynamics, facilitating both analysis and simulation \cite{jalil2013,Schwabedal-Neiman-Shilnikov-14,lodi2019design}. 

As FTM synapses do not have any temporal dynamics, the FTM paradigm does not suit the kinetics of slow synapses. When temporal dynamics of synapses become pivotal in neural circuitry, the lack of dynamics in the FTM becomes a critical limitation for modeling. This is the case where the synaptic strength changes dynamically with time and can sum up at higher ranges of spike frequency in pre-synaptic neurons to cause substantial summation in post-synaptic neurons. It was established in Refs.\cite{melibe,sakurai2016central} that slow rhythmic oscillations in neuronal circuits in sea slug CPGs require summation. Whenever the synaptic activation changes gradually, the time-varying dynamics governing the synaptic current in Eq.~(\ref{Eqsyn}) is to be described by an ODE or an ODE system.  \\ 

\paragraph{$\alpha$-synapse.} Adopting the phenomenological approach taken by Hodgkin and Huxley, Wang and Rinzel~\cite{Wang-Rinzel-92} proposed what is now commonly referred to as an $\alpha$-synapse. Its activation is meant to mimic the profile of an $\alpha$-function given by $t^p\, e^{-t}$, where $p$ is a positive integer. The idea of such $\alpha$-synapses is rooted in the pioneering computational work by W.~Rall\cite{Rall67}, who studied and modeled various aspects of synaptic potentials.
 
The dynamic equation describing the rate of change of the $S(t)$-variable in the $\alpha$-synapse is given by
\begin{equation}
S'(t) = \alpha (1-S) f_{\infty}(V_{pre}) - \beta S \label{Salpha}
\end{equation}
with some positive  $\alpha$ and $\beta$ constants. So, when $V(t)$ is below the synaptic threshold $\Theta_{syn}$ and hence $f_{\infty=0}$, then $S(t)$ exponentially decreases to zero as $e^{-\beta t}$. During an action potential as long as $V(t) \ge \Theta_{syn}$, then $S(t)$  is approaching the equilibrium state ($\alpha/(\alpha+\beta$)) exponentially fast as $e^{-(\alpha+\beta)t}$. Note that Eq.~(\ref{Salpha}) is sometimes referred to as an $\alpha-\beta$-synapse with the first-order kinetics \cite{destexhe1994synthesis}.

\begin{figure*}[!ht]
\centering
\includegraphics[width=0.85\textwidth]{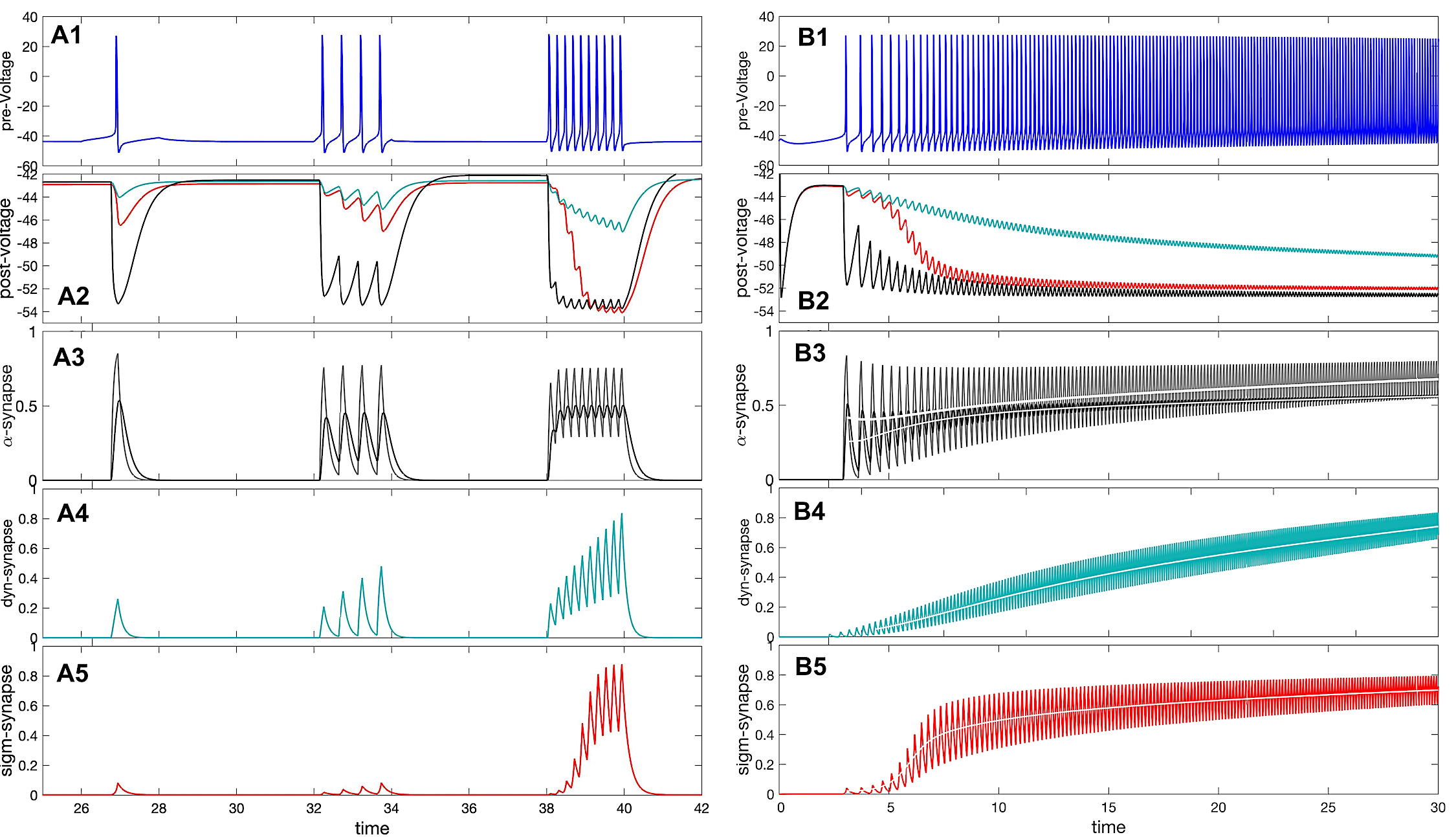}
\centering
\caption{Panel~A1: voltage trace of the pre-synaptic (quiescent) interneuron model receiving depolarizing (positive) square pulses of increasing amplitude. (A2) IPSP on the voltage trace of the post-synaptic (quiescent) interneuron model receiving inhibitory (hyper-polarizing) synaptic currents from the pre-synaptic interneuron in (A1) modeled using the $\alpha$-synapse of the first-order kinetics (A3), the slow dynamic (A4) and logistic (A5) synapses. (A3) The probability/rate of the neurotransmitter release in the $\alpha$-synapse using the 1st order (black line) with $\alpha=0.01$ and $\beta=0.008$, and 2nd order (cyan line) kinetics in response to a single spike and spike trains in the trace shown in Panel~A1. Panels~A4--A5 demonstrate stronger responses or accumulation in the dynamic ($\tau_{M}=1200$) and logistic synapses. (B1) Voltage trace shows a gradual adaptation/transition (due to slow $[\rm Ca]$-dynamics) of the pre-synaptic interneuron transitioning from an initially quiescent state  to its native tonic-spiking activity with a high frequency at $\Delta_{\rm Ca}=-70$mV. (B2) Voltage trace revealing responses of the quiescent post-synaptic interneuron receiving synaptic inhibitory current from the pre-synaptic interneuron (Panel~B1) modeled using the $\alpha$-synapse (B3), the dynamic (B4) and logistic (B5). Panels~B3--B5: increasing rate of the neurotransmitter release in the $\alpha$-, dynamical and logistic synapses correlating with the higher spike frequency in the pre-synaptic interneuron in Panel~B1. The white lines show the corresponding moving average.}
\label{fig:five_panels}
\end{figure*}

If $\alpha$ and $\beta$ values are on the same order of magnitude, say, 0.1, then the $\alpha$-synapse is as fast as an FTM \cite{Jalil-Belykh-Shilnikov-12, Alacam2015,  baruzzi2021towards, baruzzi2020generalized}. Decreasing $\beta$ by one order of magnitude $\sim 0.01$ makes the synapse sufficiently slow so that it can accumulate and demonstrate the pronounced summation with monotonically increasing $S(t)$ on average, see Fig.~\ref{fig:five_panels}A3 and B3 (grey lines), in response to trains of fast spikes in the pre-synaptic neuron. Typical values for the $\alpha$-synapses in the swim CPGs are $\alpha = 0.01$ and $\beta \in \left[0.001, 0.0005\right]$ for slow synapses, and $\beta \in \left[0.05, 0.1\right]$ for fast ones.

The dynamics of the $\alpha$-synapse can be further enhanced by adding higher order synaptic kinetics that is modeled by an ODE system with the feed-forward structure of equations like Eq.~(\ref{Salpha}) \cite{destexhe1998kinetic,DESTEXHE-CONTRERAS-SEJNOWSKI-STERIADE-94, Golomb96}: 
 \begin{align}\label{2nd}
 S_1^\prime(t) &=\alpha_1 (1-S_1) f_{\infty}(V_{pre}) - \beta_1 S_1,\\
 S_2^\prime(t) &=\alpha_2 (1-S_2)~S_1 ~~~~-~~~~~~~ \beta_2 S_2, \\
 S_3^\prime(t) &=\alpha_3 (1-S_3)~S_{2} ~~~~-~~~~~~~ \beta_3 S_3, \quad \mbox{and so on,} 
 \end{align}
with same or different time constants $\alpha_i$ and $\beta_i$ in each equation. In the case of same time constants, the synapse model of higher kinetics creates a smoothing effect and dampens the effect of individual spikes within a burst in the pre-synaptic neuron, see Figs.~\ref{fig:five_panels}A3 and B3 (black lines).

\paragraph{Dynamic synapse.}
Regarding synaptic plasticity, the accumulation of the slow $\alpha$ synapse is a poor model because it is not compatible with pronounced postsynaptic potentials. One approach to creating facilitated synapses is to introduce a modulating variable $M$ which operates on a different timescale from $S$. This approach was originally introduced to model a spike-mediated synaptic current ~\cite{Hill2001} 
\begin{equation}
\quad I_{syn} = g_{syn}\, S(t) \, M(t) \, (V_{post}(t)-E_{rev}), \label{dyn1}
\end{equation}
where $S(t)$ can be borrowed from the  fast $\alpha$-synapse~(Eq.\ref{Salpha}), or from the FTM-synapse~(\ref{ftm}), while the rate of change of the $M(t)$-variable is supposed to be quite slow  
\begin{equation}
M'(t) = \frac {f_{\infty}(V_{pre}) - M}{\tau_{M}} \label{dyn2}
\end{equation}
with a large time constant $\tau_M \sim 10^3$ or greater. We will discuss the temporal characteristics  of the dynamic synapse below. \\ 
 
We also briefly mention an additional modeling technique to alter the temporal characteristics of synaptic current while leaving dynamics of the synaptic probability $S(t)$ intact. The idea, which was also borrowed from the HH-formalism and originally intended for calibration of conductance values, is to use higher powers of $S$ in the synaptic current equation $g \cdot S^p (V-E_{rev})$, $p=2,3...$. The objective is to reshape the ascending concave-down course of $S(t)$ at an initial state to a concave-up one with a following inflection point in the time-projection due to the $S^p(t)$-term ($0 \le  S(t) \le 1$). This modification is supposed to cause initial delays and result in a less rapid/steep build-up in such a slow synapse.       

\begin{figure}[!ht]
\centering
\includegraphics[width=0.4\textwidth]{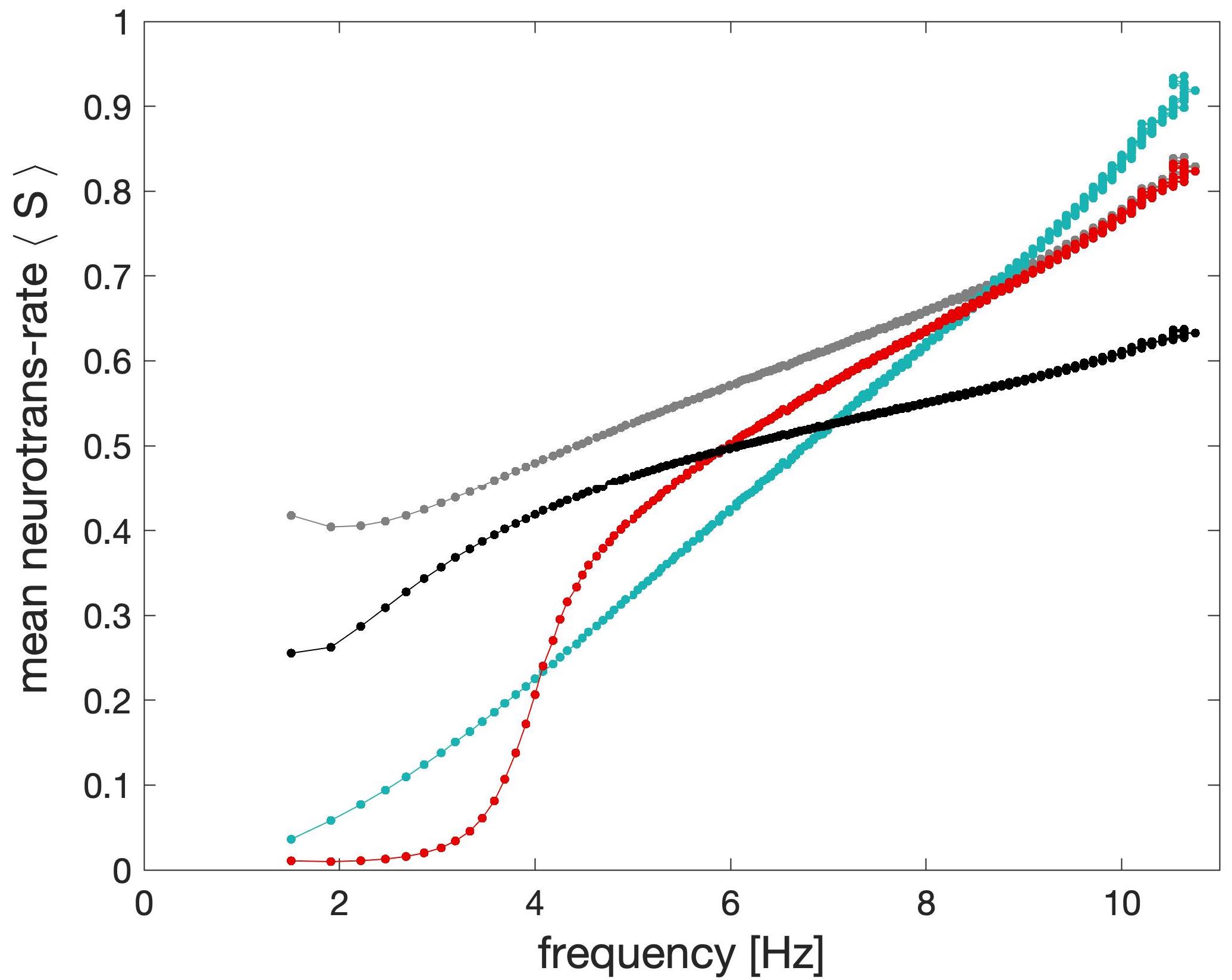}
\caption{Adopted from the moving average data in  Figs.~\ref{fig:five_panels}B3-B5: the comparison of the characteristic shapes of the {\em average} synaptic probabilities or neurotransmitter release rates $\langle S \rangle$ in the models of the $\alpha$-synapse with the 1st (grey dots) and 2nd order (black dots) kinetics, the dynamic (cyan dots) and the logistic synapses (red dots)  plotted against the spike frequency in the pre-synaptic neuron. The key feature of the logistic synapse is the inflection point in its characteristics.}
\label{fig:freq}
\end{figure} 
~\\
\paragraph{Logistic synapse.} Most slow synapses act through complex sequences of nonlinear interactions including secondary messenger cascades, which can delay the neurotransmitter release in pre-synaptic neurons or their binding with neurotransmitter receptors in post-synaptic ones. In such cases it is sometimes desirable to model a delay or some low spike frequency filter without going explicitly modeling every stage of postsynaptic intracellular signal transduction in terms of state variables. For this purpose, we introduce a new synapse model, which we call the logistic synapse, named after its similarity to the logistic model of population growth, where the probability $S(t)$ is governed by a single ODE:  
\begin{equation} \label{sigmoid}
S'(t) = \alpha\, S\,(1-S)\, f_{\infty}(V_{pre}) - \beta (S-S_0), \\
\end{equation} 
where a constant $0 \le S_0 \le 10^{-3}$ can be viewed as some spontaneous neurotransmitter release rate on average in the pre-synaptic neuron. The key feature of this modeling approach is the term $S\,(1-S)$  which provides (i) means to control a latency of the synapse in the initial ascending phase of summation, and hence (ii) a logistic or sigmoid-like shape of the $S(t)$-variable (see in Figs.~\ref{fig:five_panels}A5 and B5) determining the proportion of open channels. With this logistic model, we obtain a desired nonlinear dependence of the strength of the synaptic current on the spike frequency in the pre-synaptic neurons to match the experimental studies on the swim CPGs. In addition, with the logistic synapse acting as a high-pass filter we can further explore the role of such time-varying synaptic coupling on rhythmogenesis and its robustness in 2-cell building blocks of two types considered below.

\begin{itemize}
\item In modeling specific synaptic currents originating in specific pre-synaptic neurons, the relative values of the  $\alpha$ and $\beta$-parameter in the synaptic equations should be calibrated accordingly given the chosen synaptic threshold and the spike duration (unless unusually longer then 3ms), all matching the specific range of the spike frequencies in modeled neurons in each particular case to achieve a biologically plausible outcome. Once the range of these synaptic parameters is estimated, the strength of the synaptic current can be adjusted by varying the maximal conductance $g_{syn}$ (in nS) in targeted post-synaptic neurons, as well as by appropriately setting the synaptic reversal potential $E_{rev}$.  
\end{itemize} 

\subsection{Dynamical properties of slow synapses}
The key dynamical properties of the synapse models presented above are pinpointed in Fig.~\ref{fig:five_panels}. Its Panel~A1 shows the voltage trace of the quiescent pre-synaptic neuron model responding to the injection of three increasing depolarized pulses of the external current. Its Panel~A2 shows the inhibitory post-synaptic potentiation (IPSP) in the voltage trace of the post-synaptic interneuron responding to three inhibitory synaptic perturbations due to the spike trains in the pre-synaptic neuron (A1). These currents correspond to the $\alpha$-synapse of the first order kinetics (dark grey line), the dynamic synapse (cyan line with amplitude scaled down) and the logistic synapse (red line). In panel~A3 the varying neurotransmitter release rate $S(t)$ by the $\alpha$-synapse model of the first order (grey line) and second order (black line) kinetics (Eqs.~(\ref{2nd})) after $V_{pre}(t)\ge \Theta_{syn}$, here -40mV. Panel~A4 shows the corresponding plot of increasing rates $S(t)$ for the dynamic synapse given by (Eqs.~(\ref{dyn1})--(\ref{dyn2}). Panel~A5 shows the release rates $S(t)$ differentiated by the logistic synapse given by Eq.~(\ref{sigmoid}) which are sensitive to the spike frequency in the pre-synaptic neuron.  Note that unlike the fast $\alpha$-synapse that becomes quickly maximized and reaches saturation quickly, the strength of the slow dynamical and logistic synapses varies more gradually and depends explicitly on the spike frequency and the duration of bursts in the pre-synaptic neuron.

The $\alpha$-synapse can summate but cannot potentiate in the sense that it can warrant a slow buildup but cannot produce a sequence of postsynaptic potentials of an increasing amplitude. This is because when $S(t)$ accumulates over a spike train duration, its local minima drift away from zero so that the synaptic current, $I_{syn}$ is not significantly changed. The dynamic synapse allows for summation but does not provide an inflection point in its frequency response characteristics, and thus is an insufficient  high pass filter. Exclusively the logistic synapse can provide both a controlled delay and filter with efficiency.

The dependence of the $S(t)$ on the spike frequency in the pre-synaptic neuron is further revealed in Figs.~\ref{fig:five_panels}B1--B5. Panel~B1 depicts the spiking voltage trace of the pre-synaptic interneuron transitioning from its initial quiescent state (selected by $[{\rm Ca}] \sim 1.1$ and above) to the tonic-spiking activity at $\Delta_{\rm Ca}=-60$mV. Panel~B2 shows the hyper-polarizing effect on the post-synaptic neuron due to the growing inhibitory current produced by the $\alpha$-synapse (black line), the dynamic synapse (cyan line, and the logistic synapse (red line). Panels~ B3, B4 and B5 illustrate the different growth rates of the neurotransmitter release probability $S(t)$ through the $\alpha$-synapse model of the 1st- and 2nd-order kinetics (black and grey lines, resp.), as well as the dynamic (cyan) and logistic (red line) synapse models. Here the white lines represent the moving median $\langle S(t) \rangle$-values for all four synapse models, raising over time with the increasing spike frequency in the pre-synaptic neuron. One can see from these panels that in the case of the $\alpha$-synapse (B3), the probability of the opening channels quickly reaching half of their capacity as soon as the interneuron transitions from the quiescent state to spiking activity. After that, the rate increases slowly with the increase of the spike frequency.  While the monotone dependence is preserved in the dynamic synapse, its average rate $\langle S \rangle$ grows almost linearly as the spike frequency is increased (panel B4). In contrast, the dependence of $\langle S \rangle$ on the spike frequency in the logistic synapse is expectedly nonlinear, and of the sigmoidal shape (panel B5).

The moving averages $\langle S(t) \rangle$ shown as the white lines in Figs.~\ref{fig:five_panels}B3-B5, are further utilized in Fig.\ref{fig:freq} to determine the dependence of the strength of the synapse models on the spike frequency in the pre-synaptic neurons in the long run, skipping initial transients. One can see from this diagram that in the case of the $\alpha$-synapse of the first (grey) and second (black) order kinetics there is a threshold around 1.5--2Hz, after which its strength escalates nearly instantaneously to increase slowly with the higher spike frequency.  Observe that the dependence of the strength of the dynamic synapses on the spike frequency is nearly monotonically linear, likely due to its second slowly varying constituent term given by Eq.~(\ref{dyn2}).

The main feature of the proposed logistic synapse, which is also reported in the experimental studies on the sea slugs is that the its strength remains weak as long as the spike frequency in the neurons stays below 3--4Hz (see also Fig.~\ref{fig:five_panels}A5). With the higher spike frequency, its efficiency and strength rapidly grow within the spike range 3--6Hz and keeps increasing at the slower rate at greater frequencies.   

We note that by increasing the corresponding $\beta$-value, while keeping $\alpha$ same in the synapse model, one can rectify the dependence of the mean synaptic probability $\langle S \rangle$ with a flat saturation in the characteristics at greater spike frequencies. Such a saturation may occur even before the neuron model becomes the depolarized quiescent state at greater positive values of the external current applied or due to small negative values of $\Delta_{\rm Ca}$. In the terminology of bifurcations, at this transition the corresponding periodic orbit on the critical manifold $M_{\rm po}$ in the phase space of the neuron model shrinks in amplitude and collapses into the depolarized steady state through a sub-critical Andronov-Hopf bifurcation, see Fig.~\ref{fig:Plant_burster}A.

\begin{figure}[!ht]
\centering
\includegraphics[width=0.48\textwidth]{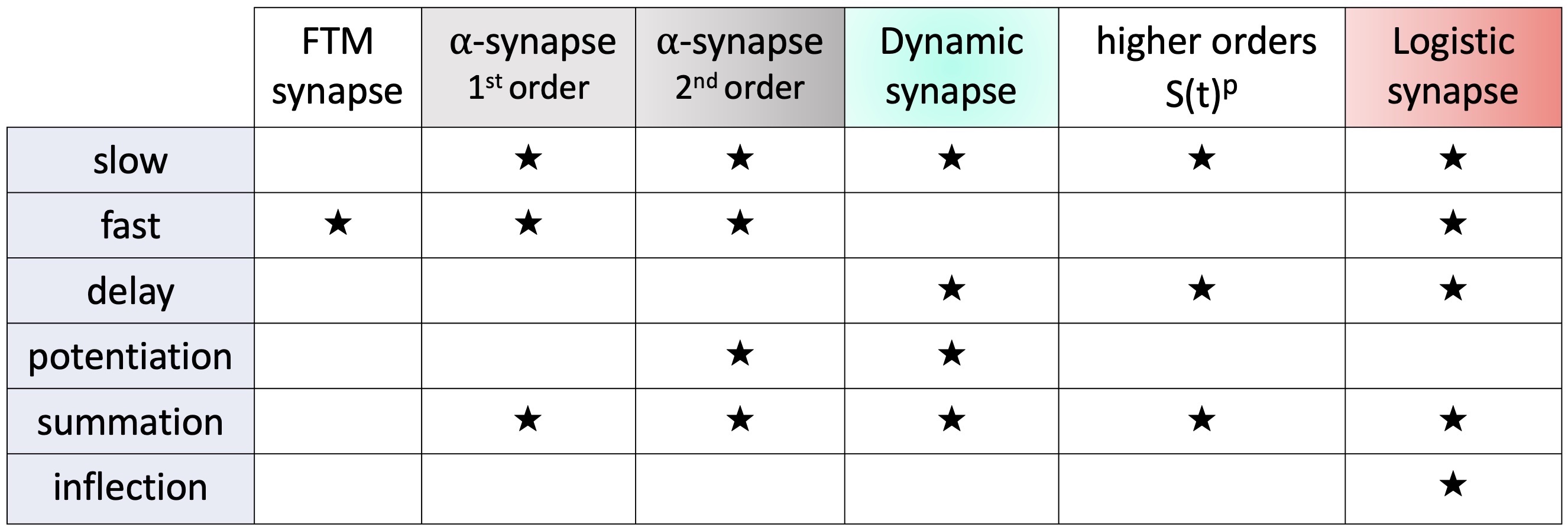}
\caption{Key properties of the synapse models.}
\label{fig:table}
\end{figure}

The following table~\ref{fig:table} summarizes the dynamical and temporal characteristics of the synaptic models discussed above. 

In this table, {\em inflection} refers to the concavity change point in the sigmoid characteristics of the logistic synapse in Fig.~\ref{fig:table}. Here, summation refers to a nearly monotone, solid buildup of I/EPSPs in a voltage trace of the post-synaptic neuron caused and aligned with the spike train in the pre-synaptic one. In contrast, in the case of potentiation the size of I/EPSPs increases with each sequential spike in the pre-synaptic neurons, followed by a reset to a base-line between sequential spikes. The potentiation typically occurs in a synapse model described by an ODE system  with diverse fast and slow time-scales, such as Eqs.~(\ref{2nd})-(18) or Eqs.~(\ref{dyn1})-(\ref{dyn2})  due to distinct $\alpha$-values on different magnitude orders, say $\alpha_1=1$ and $\alpha_2=0.1$, or $\alpha_1=1$ and a large $\tau_m$-constant, respectively. Note that with a higher spike frequency, potentiation may likely morph into summation.           

\section{Building blocks of rhythmic neural circuits} \label{sec:4}

We will consider two kinds of pair-wise neural circuits, which represent small building blocks of large networks. The first pair-wise circuit, initially described in the classic Brown's paper \cite{brown} is called a half-center oscillator (HCO). A typical HCO denotes a pair of neurons reciprocally coupled by inhibitory synapses to burst in alternation. In theory, HCO neurons are supposed {\em not} to burst endogenously or after being injected with a constant external current; i.e., they cannot be latent bursters either \cite{Alacam2015}. This suggests the exclusion of any hysteresis due to slow variable(s) in the adopted model and similar models of individual neurons.

There are several loosely defined mechanisms underlying bursting in small networks such as HCOs, see Refs.\cite{Perkel-Mulloney-74, Wang-Rinzel-92,SKM94, SZVC99, Kopell-Ermentrout-02, Angstadt-Grassmann-Theriault-Levasseur-05, Matveev-Bose-Nadim-Farzan-07, Shilnikov2008b, Silverston2009,Daun-Rubin-Rybak-09, Jalil-Belykh-Shilnikov-10,nagornov2016mixed} and references therein. One mechanism is the {\em post inhibitory rebound} (PIR) discussed previously and illustrated in Fig.~\ref{fig:two_panels}. The PIR typically occurs in a quiescent neuron when it becomes rapidly depolarized and produces  a spike train after its release from hyper-polarization forced by an inhibitory current. The HCO neurons rebound sequentially to produce alternate bursting, where either active pre-synaptic neuron inhibits its post-synaptic counterpart and vice versa. Another mechanism is called an {\em escape}. In this context, the escape refers to the configuration where the active pre-synaptic neuron producing tonically spikes is ``over-thrown'' into the inactive quiescent phase by the inhibition from the counterpart neuron after its fast transition (including PIR) from the inactive quiescent phase into the tonic-spiking phase. A third mechanism is referred to as a {\em release} from inhibition in the hyperpolarized post-synaptic neuron after the pre-synaptic neuron ends its active tonic-spiking phase. Note that mechanisms all three mechanisms can be often combined to generate and maintain network bursting in the HCO.   

In what follows we would like to enhance the HCO framework further to classify network-level rhythm generating mechanisms by redefining some useful concepts. We say {\em emergent} bursting in networks occurs provided that i) neither neuron bursts in isolation and ii) it is solely due to transient dynamics, cellular and synaptic. Such transient dynamics are mainly maintained by slowly varying synaptic kinetics in respond to the spike frequency adaptation modulated by the slow $[\rm Ca]$-circulation in the given neuron model. This leads us to introduce a new concept of a {\em network hysteresis}. This phenomenon emerges due to cyclic overlapping emerging between branches of the tonic-spiking manifold $M_{\rm po}$ and the quiescent manifold $M_{\rm eq}$, which becomes distorted in the post-synaptic neuron by the synaptic inhibition enforced by the pre-synaptic one. Alternatively, in the $(\rm Ca^{2+},x)$-projection on the slow subsystem, the overlap emerges between the average $\langle x \rangle$-nullcline in the pre-synaptic neuron and the nullcline  $x^\prime=0$ in the inhibited post-synaptic neuron, see Figs.~\ref{fig:two_panels}A3 and B3. We argue that i) the spike frequency adaptation mediating ii) slowly graduating synaptic coupling are the pivotal elements that provide stability, flexibility, and resilience to parameter variations and perturbations in oscillatory neural circuits. 

Below we will demonstrate two different mechanisms of rhythm generation in such pair-wise networks. The first example is emergent bursting in a reciprocally inhibitory network, reminiscent of Brown's original idea. Its mechanism is based on the concept of the network hysteresis. The second neural pair consists of excitatory--inhibitory neurons where the tonic-spiking neuron provides a forward excitatory drive to the quiescent or less active neuron that subsequently provides a slowly building inhibition in the backward loop. This excitatory--inhibitory module is reminiscent of the {\em predator--prey} concept of a biological oscillatory with a distinctive 1/4 phase-lag,  whereas HCO oscillations are somewhat analogous to the visual phenomenon known as the {\em binocular rivalry}, when the spectator can observe two cyclically alternating images in a single picture. In this case oscillations occur in anti-phase with a 1/2 phase-lag or alternate with intervals close to half-period of the network.

\section{Half-center oscillator} \label{sec:5}

\begin{figure*}[!ht]
\centering
\includegraphics[width=1.0\textwidth]{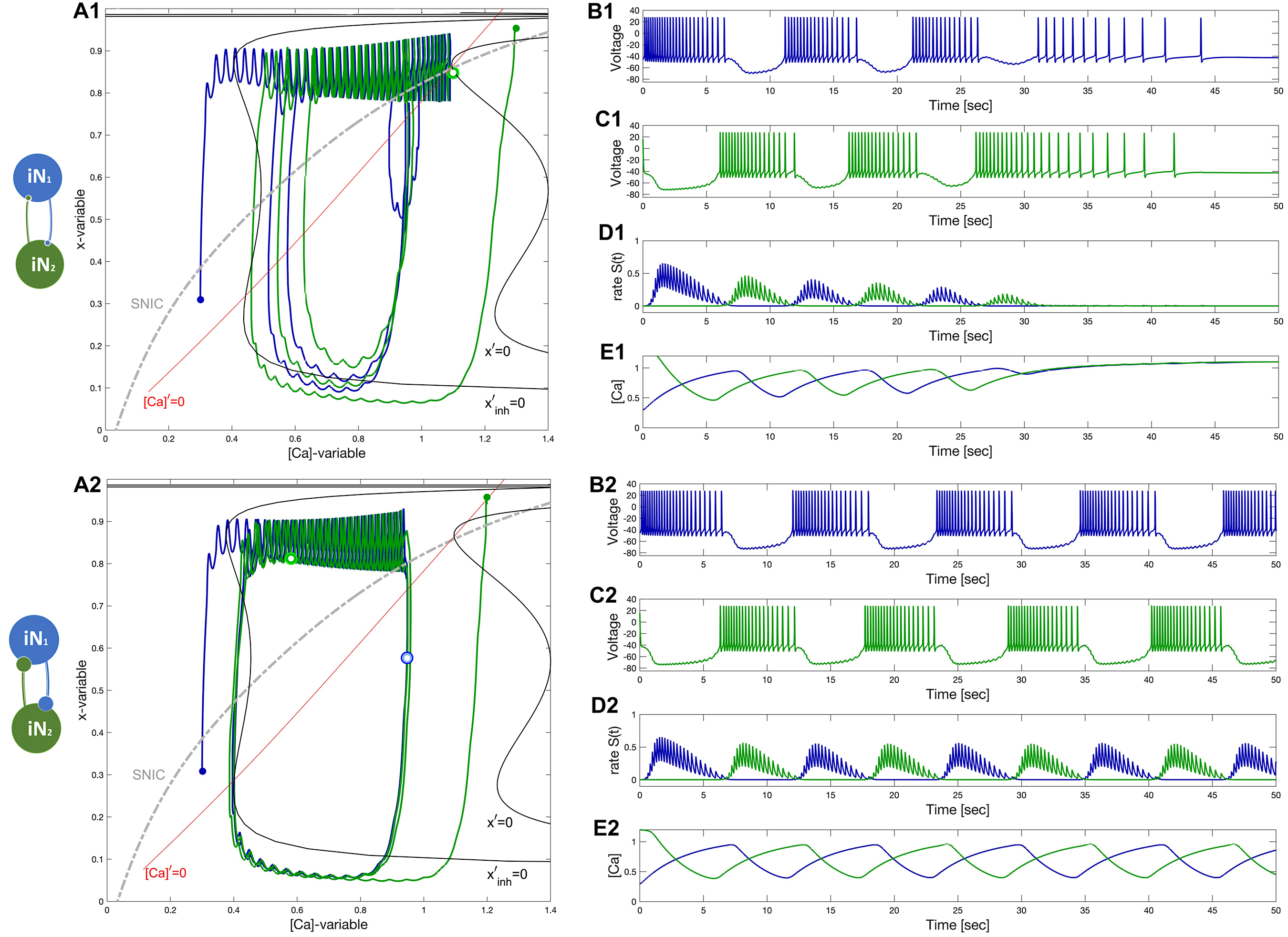}
\caption{A HCO-network constituted of two quiescent interneurons, iN\textsubscript{1} (blue) and iN\textsubscript{2} (green) at $\Delta_{\rm Ca}=-30$mV, coupled reciprocally by the inhibitory (denoted by $\bullet$ ) logistic synapses. Schematic diagram (left). Panels A1--E1: insufficient inhibition ($S(t)$-traces in Panel~D1) and/or improper phase-lag (small green and blue dots) between initial phases of the HCO interneurons does not let emerging network-bursting keep the initial momentum and oscillations seize and converge to a quiescent state (voltage traces in Panels B1 and C1), which is located at the crossing of the nullclines $x'=0$ and $[\rm Ca]'=0$ (red and black lines, resp.) below the SNIC-curve (as a dotted grey line)  in the $([\rm Ca],x)$-phase plane in Panel~A1.    
Panels A2--E2: increasing the reciprocal inhibition strength leads to the onset of self-sustained emergent bursting in the HCO.  HCO bursting is associated with a stable cycle in the $([\rm Ca],x)$-phase plane shown  in Panel~A2, which occurs due to the emergent network hysteresis in the driven neuron during the slow hyperpolarized phases near the forced nullcline $x_{\rm ing}^\prime=0$. This HCO is {\em bistable}: bringing initial states of the neurons closer will lead to network oscillations damping with time as in the previous case.}
\label{fig:hco1}
\end{figure*}

A reciprocally inhibitory HCO can be composed of either quiescent or tonic spiking neurons, or their combinations. Let us first discuss the case where the HCO is made of two quiescent interneurons. Its two snapshots are shown in Fig.~\ref{fig:hco1}. It is apparent from the traces in  Fig.~\ref{fig:hco1}B1 and C1 that when the reciprocal inhibition is insufficient or the initial phases of the constituent neurons are not diverse (too close) enough, or both,  then emerging network bursting cannot keep up its momentum and the oscillations decays and converge to the steady state originally observed at $\Delta_{\rm Ca}=-30$mV. The corresponding equilibrium state in the projection to the $([\rm Ca],x)$ phase plane is located at the intersection of the calcium nullcline $[\rm Ca]^\prime=0$ (red line) and the $\Sigma$-shaped equilibrium state nullclines: unperturbed $x^\prime=0$ and inhibition-perturbed $x^\prime_{\rm inh}=0$, shown in  Figs.~\ref{fig:hco1}A1 and A2. Note that the superimposed $x$-nullclines: $x^\prime=0$ and $x^\prime_{\rm inh}=0$, correspond to the cases where i) there is no interaction between the interneurons located at the equilibrium state (double dots) and ii) to the perturbed nullcline respectively, which is shifted to the left by the inhibition. Note that the position and shape of the latter are not constant and transition toward the unperturbed nullcline $x^\prime=0$ as the inhibition decreases, as can be seen in the traces for the synaptic probability $S(t)$ in Fig.~\ref{fig:hco1}D1. We indicate too that the intersection point of the nullcline $[\rm Ca]^\prime=0$ with the unperturbed one $x^\prime=0$ is located on its stable section and hence corresponds to the stable equilibrium state for the hyperpolarized quiescence. Meanwhile, such an intersection point on the perturbed nullcline $x^\prime_{\rm inh}=0$ is located above the bottom knee point and hence correspond to the unstable equilibrium state of the network. 

\begin{figure*}[!t]
\centering
\includegraphics[width=1.0\textwidth]{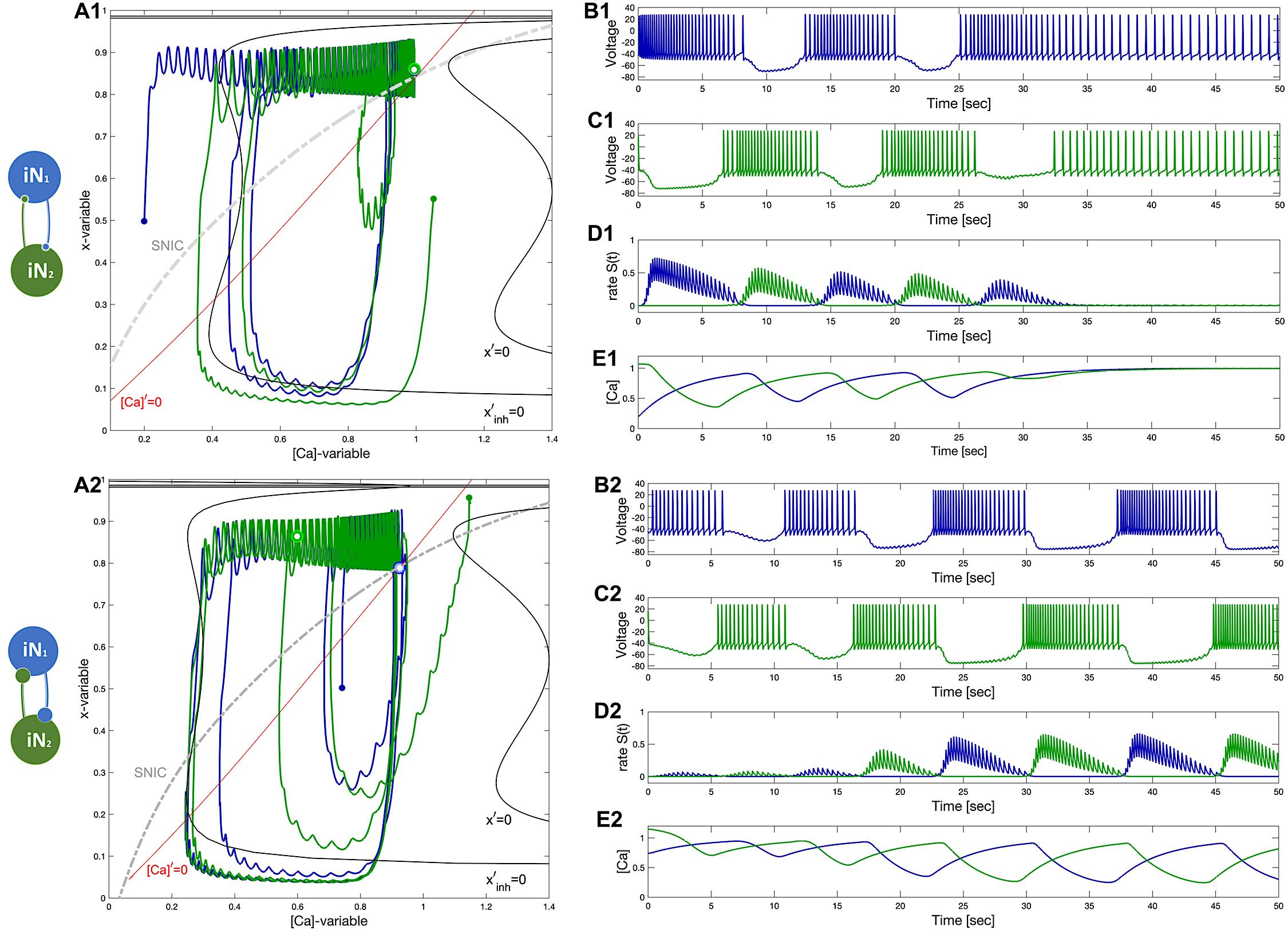}
\caption{A HCO-network constituted of two tonic-spiking neurons at $\Delta_{\rm Ca}=-40$mV. Panels A1--E1: insufficient reciprocal inhibition and/or phase-lag between initial phases cause network bursting to weaken and make both neurons return to natively tonic-spiking activity, which can be well-observed in $[\rm Ca]$-traces (Panel~E1) converging to some fixed value corresponding to two round periodic orbits indicated by a double dot above the SNIC-curve near the (red) nullcline $[\rm Ca]'=0$ in the phase plane in Panel~A1. Panels A2--E2: changing initial phases and/or increasing the reciprocal inhibition leads to the onset of network bursting, corresponding to a stable limit cycle in the $([\rm Ca],x)$-phase plane shown in panel~A2. This HCO is {\em bistable}: decreasing inhibition or/and initial phase-lag will lead to similar damping oscillations as in the previous case. }
\label{fig:hco2}
\end{figure*}

Increasing the inhibition makes emerging bursting develop into the robust, self-sustained rhythm in the HCO  as one can observe from  Figs.~\ref{fig:hco1}A2, B2, and C2. Note that with the given coupling strength, the oscillations emerge and sustain also because of properly chosen initial phases defined by diverse $[\rm Ca]$-values (indicated by small green and blue dots in Fig.~\ref{fig:hco1}A2) in the constituent neurons. For other values, the network oscillations would have ultimately converged to the stable equilibrium state as it in the previous case. Let us reiterate that the stable limit cycle observed in the $([\rm Ca],x)$ phase plane, and corresponding to the self-sustained anti-phase bursting oscillations in the HCO, encloses the unstable network equilibrium state located at the crossing point of both nullclines, just like it is supposed to be by virtue of theory of systems in a plane.      

To this point, we can write down some preliminary observations: 
\begin{itemize}
\item[i.] the HCO is bi-stable; 
\item[ii.] the inhibition must be sufficient to warrant the emergence of stable network bursting;
\item[iii.] the balance of initial phases must be also fulfilled to initiate and maintain network anti-phase bursting.    
\end{itemize}  

\begin{figure*}[!t]
\centering
\includegraphics[width=1.0\textwidth]{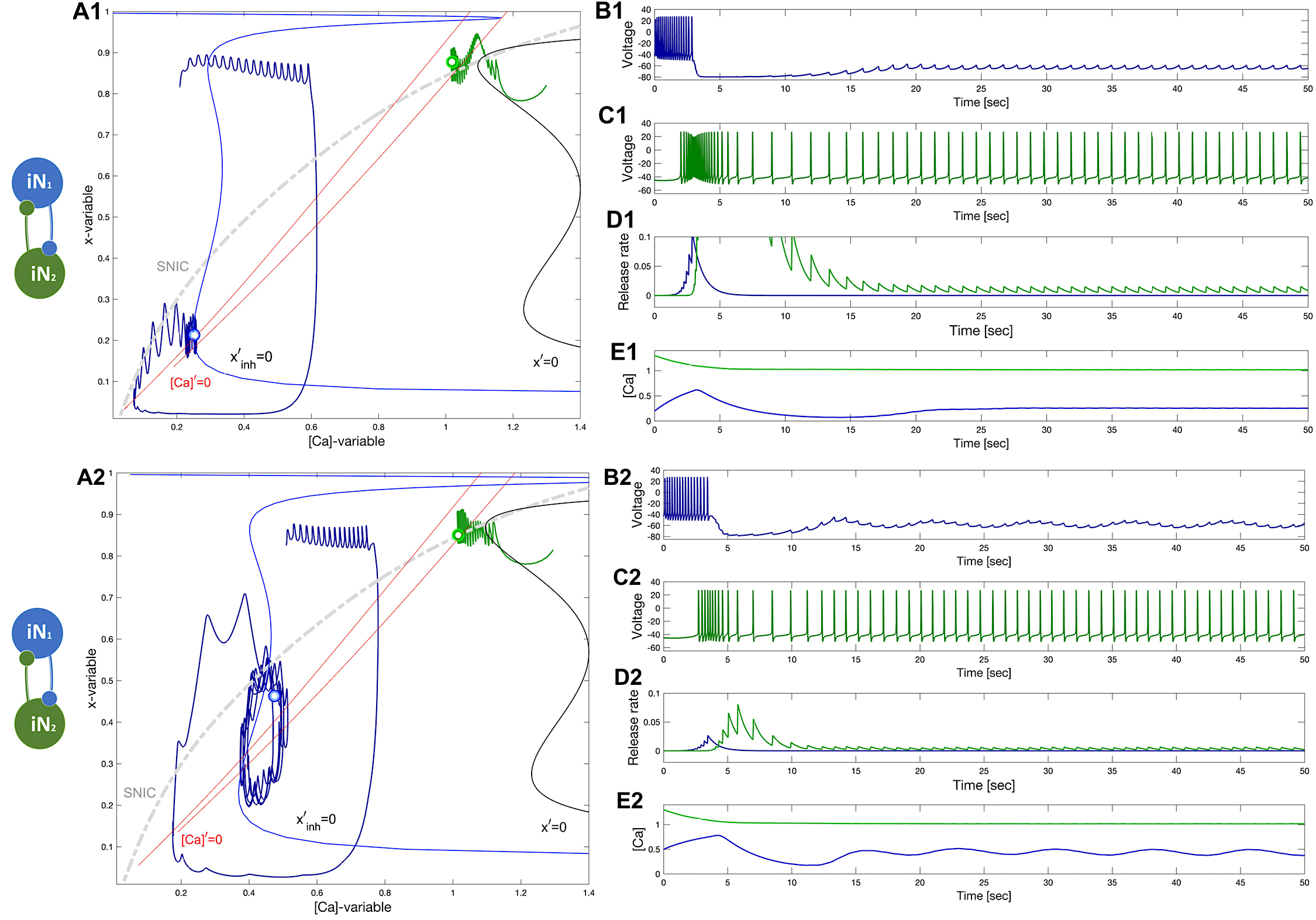}
\caption{Illustration of the concept ``winner takes all'' in the HCO network. Panels A1--E1: The initial positions place the green neuron at $\Delta_{\rm Ca}=-40$mV atop so that its strong and lasting inhibition forces the blue counterpart neuron at $\Delta_{\rm Ca}=-50$mV to stay at a hyperpolarized state near the intersection of the corresponding nullclines $[\rm Ca]^\prime=0$ and $x^\prime=0$ (next to its bottom knee-point) and under the SNIC-curve in the phase plane as shown in panel A1. Panels A2--E2: Weakening inhibition makes the blue neuron become less hyperpolarized to produce forced sub-threshold oscillations seen in the voltage traces and corresponding to a limit cycle in the $([\rm Ca],x)$-phase plane between the two knee-points of the $\Sigma$-shaped nullcline $x^\prime=0$ as depicted in Panel~A2.
}
\label{fig:hco3}
\end{figure*}

Note that both conditions ii--iii) above are to be met: initial phases of the neurons and their coupling strength are set to move the HCO far away from its natural equilibria to make the HCO  {\em unstable} and to initiate its perpetual motions due to nonlinear reciprocation between its cellular and synaptic components. For example, to set initial phases correctly, either HCO neuron should be in the active depolarized state while the other one should initially be placed in the inactive hyperpolarized state. Another option to trigger network oscillations is to place both neurons initially at depolarized states quite far from their natural equilibria, which are either stable quiescent states or periodic orbits representing tonic-spiking activity as in our following case below. In biological swim CPGs, such initial conditions of the interneurons can be turned ``on'' by an excitatory drive from some sensory cells to trigger and initiate the locomotion.

We stress that in the given HCO configuration where both neurons ``chase'' each other to reach their stable {\em natural} equilibria, any synapse models discussed above should work reliably because the uni-directional inhibition vanishes as soon as the pre-synaptic neuron becomes quiescent and releases its counterpart with the following post-inhibitory rebounds to start a new network burst cycle. However, faster FTM and- $\alpha$-synapses will make the HCO antiphase dynamics stiffer and more sensitive to parameter selections, unlike the logistic synapses that warrant more flexibility, including asymptotical stability of the network attractor. This assertion is manifested by the convergence revealed in the traces in Figs.~\ref{fig:hco1}B1--E1 and B2--E2, is further supported the network limit cycle shown in the $([\rm Ca],x)$-phase plane in Fig.~\ref{fig:hco1}A2. It displays that there are two (blue and green) double dots representing the snapshot of the instantaneous phases of the neurons on the network limit cycle in the $([\rm Ca],x)$-phase plane, see also Fig.~\ref{fig:hco1}A2. The mutual positions of these instantaneous phases indicate that network bursts can significantly overlap thereby testifying and quantify the level of robustness of the HCO coupled with the logistic synapses, as well as its structural stability with respect to parameter variations.          

\begin{figure*}[!t]
\centering
\includegraphics[width=1.0\textwidth]{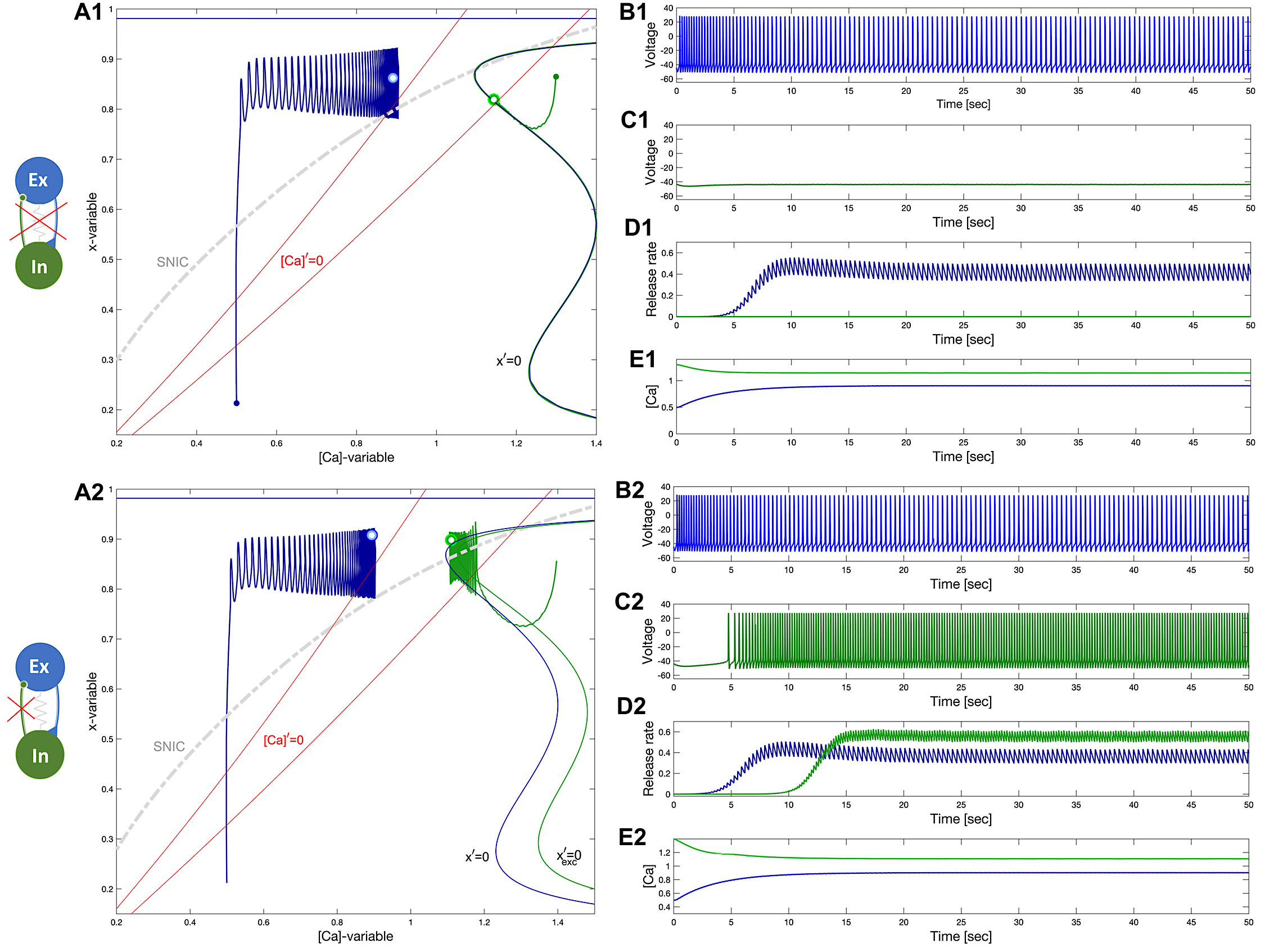}
\caption{Building an EI-module. Panels A1-E1: uncoupled excitatory tonic-spiking neuron (blue) at $\Delta_{\rm Ca}=-50$mV and inhibitory quiescent neuron (green) at $\Delta_{\rm Ca}=-20$mV. Panels A2-E2: Turning the excitatory synapse ``on'' in the blue neuron makes the inhibitory neuron spike tonically as well. (A1 and A2) red lines labeled by $[\rm Ca]^\prime=0$ standing for the calcium nullclines above/below when the $[\rm Ca]$-variable increases/decreases, the unperturbed and perturbed locations of the equilibrium states (double dots) are shown in the color-matching $\Sigma$-shaped nullclines $x^\prime=0$ and $x^\prime_{\rm exc}=0$, as well as shows the position of the periodic orbits in the $([\rm Ca],x)$-phase planes  for both neurons.}
\label{fig:fredo1}
\end{figure*}

This statement is further supported in the case where the HCO is made of two tonic-spiking neurons coupled with the logistic synapses to ensure greater flexibility in the network dynamics, see Fig.~\ref{fig:hco2}.  We should re-emphasize that the specific values of the $\alpha$ and $\beta$-constants in Eq.~(\ref{sigmoid}), including the maximal conductance $g_{syn}$, must be calibrated to match the spike frequency in the isolated neurons (at $\Delta_{\rm Ca}=-40$mV) to ensure the emergence and stability of HCO bursting. In simple terms, the level of the synaptic probability $S(t)$, and hence the inhibition force should weaken significantly with the decreasing spike frequency within bursts, see voltage traces in Fig.~\ref{fig:hco2}B1 and B2. This will guarantee that the driven neuron is realized from its forced state with a strong post-inhibitory rebound, which is determined by its $\rm Ca^{2+}$-concentration, which, in turn, is determined explicitly by the inhibition strength.       

This is fully manifested in Fig.~\ref{fig:hco2}: its panels demonstrate step-by-step the stages of network anti-phase bursting emerging asymptotically through a canard-type periodic orbit of initially small amplitude through an Andronov-Hopf bifurcation, as often occurs in slow-fast systems. Note that slowing down the gating $x$-dynamics equates its time scale with that of the $[\rm Ca]$-dynamics. Then, the emergence and onset of the stable network oscillations (and the evolution of the corresponding limit cycle) can occur more gradually and consistently with a genetic Andronov-Hopf bifurcation. 

Figures~\ref{fig:hco2}A1--E1 demonstrate that when the synaptic coupling is slightly below some bifurcation threshold, network oscillations decay exponentially with time in figures ~\ref{fig:hco2}B1-E1, while large oscillations shrink in size and  converge to the tonic-spiking (round) periodic orbit depicted in the $([\rm Ca],x)$-phase plane in Fig.~\ref{fig:hco2}A1. Its location, to the left from the fold on the nullcline $x^\prime=0$ (slightly perturbed and shifted leftwards by the reciprocal inhibition), can be evaluated by the position (white/green dot) of the nullcline $[\rm Ca]^\prime=0$  (red line).

The asymptotic stability of emergent HCO network bursting is documented in Fig.~\ref{fig:hco2}A2. It can be well-seen in the time progressions of the growing synaptic rates $S(t)$ (Fig.~\ref{fig:hco2}D2) and of the calcium concentration $[\rm Ca](t)$ (Fig.~\ref{fig:hco2}E2) increasingly oscillating in anti-phase in the HCO. Let us reiterate that even though the coupling strength, in this case, can be sufficient to sustain such network oscillating, the lack of the balance of the initial phases results in the failure as in the Fig.~\ref{fig:hco2}B1. This can be better illustrated using the $([\rm Ca],x)$-plane Fig.~\ref{fig:hco2}A2. Shifting the initial condition (blue dot) of the blue neuron to the right, to a larger initial $[\rm Ca]$-value, misbalances some necessary relationship between initial phases thereby breaking down the HCO oscillations. This means that the HCO made of tonic-spiking neurons is a bistable network with two possible states: both coupled neurons exhibit tonic-spiking activity, or generate slow network bursting.

This observation provides hints to successfully initiate network bursting: option i) either neuron is initially set in the inactive, hyperpolarized phase, while the other neuron starts being depolarized; option ii) both neurons are initially excited by an external current or synaptic pulse to trigger the inhibitory race between them, which naturally puts them in the opposite ``on'' and ``off'' conditions to develop into anti-phase bursting in the HCO. Lastly but not least, adding properly calibrated channel noise can make emergent bursting even more robust and look realistic with adequate fluctuations in burst duration and period. We note that channel noise (in the range $\le 14$Hz) can become an influential player that helps network bursting keep running in the extreme case considered below.   

\begin{figure*}[!t]
\centering
\includegraphics[width=1.0\textwidth]{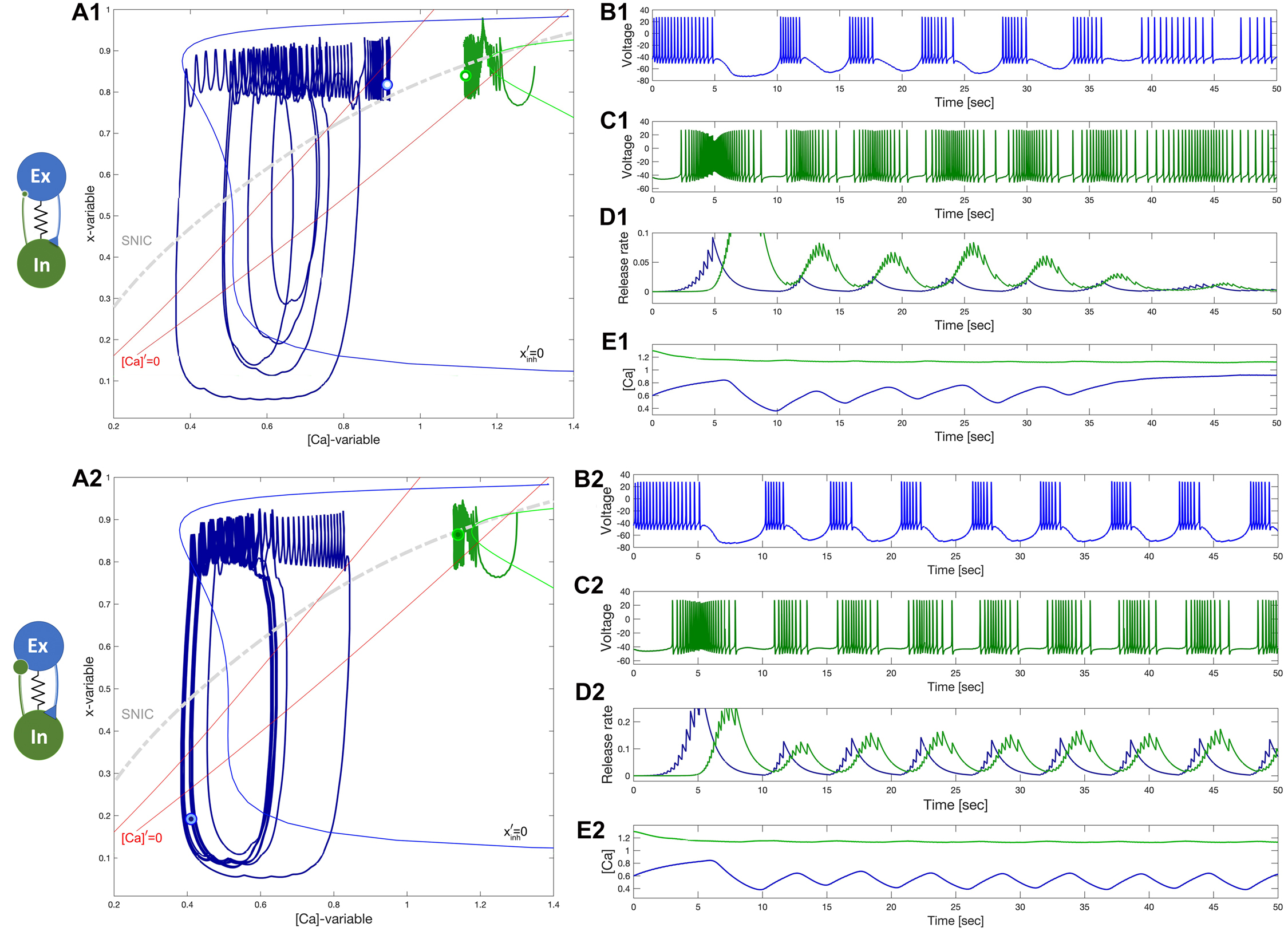}
\caption{Building an EI-module. Panels A1--E1: Turning the inhibitory synapse ``on'' makes the EI-module initiate network bursting but cannot sustain due to a lack of  balanced coupling so it breaks apart with its components returning to their natural states: a tonic-spiking neuron  (blue) at $\Delta_{\rm Ca}=-50$mV and a quiescent neuron (green) at $\Delta_{\rm Ca}=-20$mV. Panels A2--E2: increasing inhibition lets the EI-module sustain the network bursting stably as seen from the voltage and release rate $S(t)$-traces as well-seen from Panel B2-D2. Note both neurons alternate between tonic-spiking and quiescent phases demarcated by the SNIC-curve in the $([\rm Ca],x)$-phase plane in panels A1 and A2.}
\label{fig:fredo2}
\end{figure*}

\subsection{``Winner takes all''}
         
To complete the discussion on HCO-dynamics let us consider the case where {\em winner takes all}. To ensure some generality we note that variance in the spike frequency of neurons is common: so in this setup the blue neuron demonstrates a high spike frequency at $\Delta_{\rm Ca}=-50$mV, while the green neuron is less excited with a lower spike frequency at $\Delta_{\rm Ca}=-40$mV. The difference should result in uneven inhibitory outcomes. Nevertheless, as seen in Fig.~\ref{fig:hco3}A1, with given initial phases the green neuron becomes a winner that inhibits and shuts down the blue one at the forced hyperpolarized state (Fig.~\ref{fig:hco3}B1 and C1 first and second traces). This hyperpolarized state is located at the intersection of the stable (solid line) bottom section of the forced nullcline $x^\prime{\rm inh}=0$ (blue $\Sigma$-shaped curve) and the nullcline $[\rm Ca]^\prime=0$ in the $([\rm Ca],x)$-plane in Fig.~\ref{fig:hco3}A1. Such an extreme case may occur when the {\em lasting} inhibition is too strong either due to a larger $\alpha$-value or lower $\beta$-value in the logistic synapse. Furthermore, weakening the inhibition produced by the winner (green) -- the loser (blue) neuron can give rise to sub-threshold oscillations in the forced post-synaptic neuron, see voltage traces in Fig.~\ref{fig:hco3}B2. Observe from the traces that small ``vibrations'' in the transmission rate $S_{green}(t)$-precession (Fig.~\ref{fig:hco3}D1) and the IPSPs in $V_{blue}(t)$-trace (Fig.~\ref{fig:hco3}B1) are directly correlated. These sub-threshold oscillations are associated with a stable network limit cycle in the phase plane depicted in Fig.~\ref{fig:hco3}A2. The limit cycle between two bottom folds of the $\Sigma$-shaped nullcline $x^\prime=0$ is a solid replica of the classic relaxation or the FitzHugh-Nagumo oscillator. Note that with further weakening inhibition from the winner neuron, the size of the forced limit cycle in the blue (loser) neuron increases and starts crossing the SNIC-curve. Correspondingly, the amplitude of subthreshold oscillations in the blue voltage traces will also increase with the addition of a few new spikes, transforming eventually into full bursting episodes.      

We close the HCO section with the following conclusions: 

\begin{itemize}
\item HCO bursting is a network-level emergent phenomenon based on the novel notion of network hysteresis. 
\item  HCO bursting becomes self-sustained provided that both the balance of phase and  the balance of amplitudes (coupling) are fulfilled;
\item If the above conditions fail, then the HCO falls apart and converges to the natural states of its  constituent  neurons; 
\item  this implies that the HCO network is a bistable one;
\item logistic synapses make the HCO-dynamics more flexible, and less stiff as in the case where similar $\alpha$-synapses are employed; 
\item Such a  HCO can be made of both (non-identical) quiescent, or both tonic-spiking neurons, or their combinations, which warrants its broad structural stability range.      
\end{itemize}

\section{Excitatory -- inhibitory (EI) module} \label{sec:6}

\subsection{Tonically spiking and quiescent neurons}

As mentioned above, there is a parallel between the way the oscillations with a specific phase-lag or period fraction emerge in such an excitatory-inhibitory (EI) module and the classic predator-prey paradigm. Here the excitatory (blue) neuron is a ``prey'' that provides enough drive that lets the inhibitory (green) neuron, playing the role of the ``predator,'' generate tonic-spiking activity to slow down the excitation through the emerging inhibitory feedback. The inhibition then forces the excitatory neuron to transition into an inactive hyperpolarized phase during which the predator loses the excitatory drive which has to come back to its initial state release the prey from inhibition. The logistic synapses happen to be a great match for this concept as they can produce fast excitation in the forward positive loop and slowly building inhibition in the negative feedback loop.

\begin{figure*}[!t]
\centering
\includegraphics[width=1.0\textwidth]{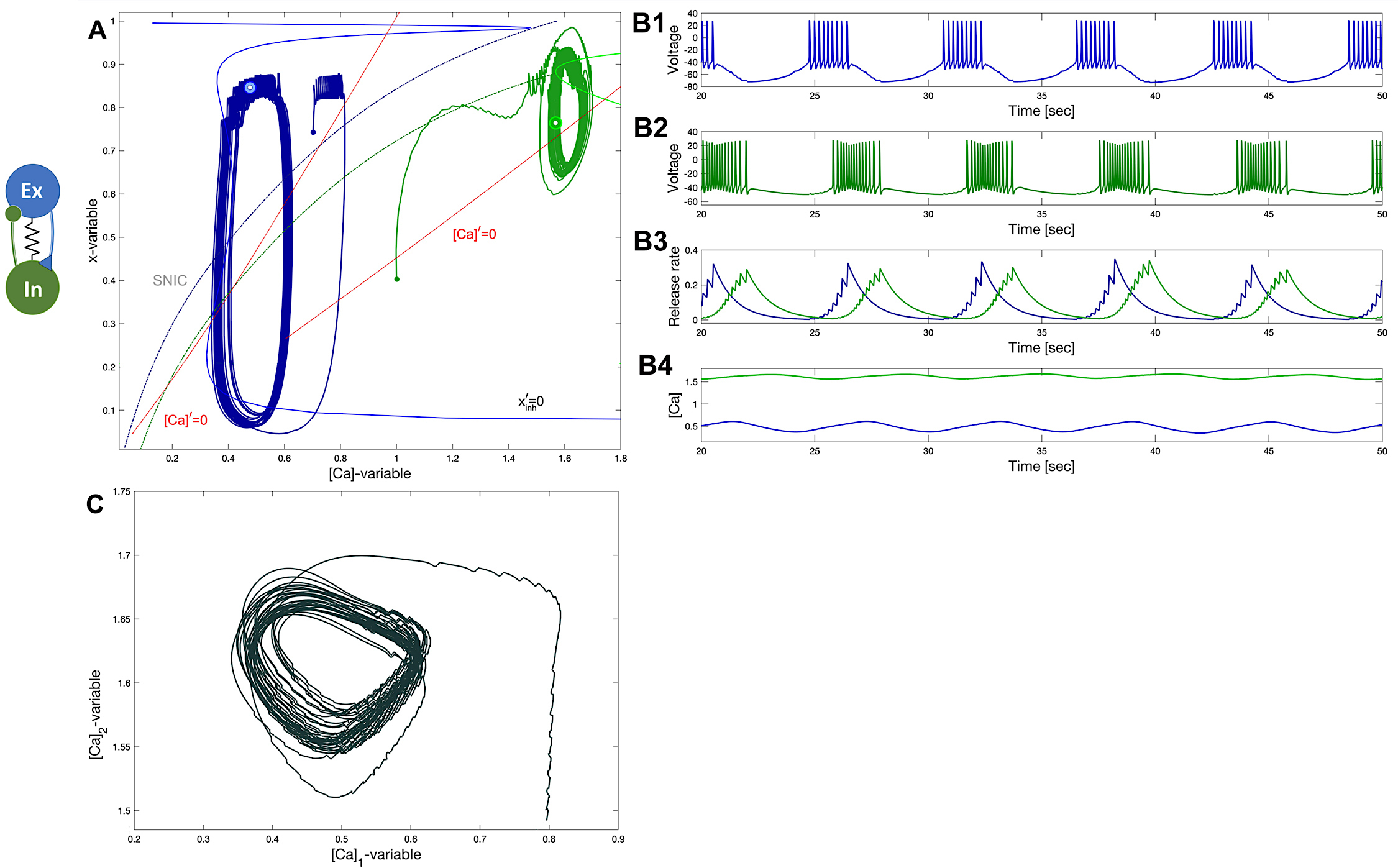}
\caption{{\it Prey-predator network.} A diverse EI-module constituted of a tonic-spiking neuron (blue) at $\Delta_{\rm Ca}=-60$mV and a quiescent neuron (green) at $\Delta_{\rm Ca}=+60$mV that reciprocally produce robust bursting in voltage traces (B) corresponding to a stable limit cycle in the $([Ca],x)$-plane in Panel~A  and $\left ([\rm Ca]_1,\, [\rm Ca]_2 \right )$-phase plane in Panel~C due to both factors: sufficiently strong coupling and a proper phase-lag $\sim 1/4$ period of oscillations between excitatory and inhibitory fluxes of neurotransmitter releases seen in Panel~B3. Decreasing the inhibitory strength makes the oscillations less pronounced. Note two phase point snapshots, blue and green dots located at 12 o'clock (above) and 9 o'clock (left) on the limit cycles lock positions on the two cycles in Panel A indicating a desired 1/4-period phase-lag between the oscillations in the excitatory and inhibitory neurons, resp.}
\label{fig:fredo3}
\end{figure*}

In this configuration, setting the reversal potential $E_{rev}=40$mV makes a synapse excitatory. In addition, there is a weak electrical synapse or a gap junction between the neurons. The gap junction was reported in the biological EI-module, which happens to be the key building block in the swim CPG of the sea slug {\em Dendronotus iris} \cite{sakurai2022bursting}. Unlike the chemical synapses, the gap junction is bi-directional, and the corresponding current is described by an additional term $I^{elec}_{syn}=g_{elec}(V_{post}-V_{pre})$, in the voltage equation~(\ref{eq:voltage}), where $g_{elec} \simeq 1 - 2 \times 10^{3}$nS. We argue that the role of the gap junction is two-fold as always and depending on its strength and neural excitability it can aid to stabilize the network oscillations, or dampen them when the balance of amplitudes (due to coupling strength) and phases are insufficient. Note the classic predator-prey setup is (energy) conservative, and hence lacks dissipation.

\begin{figure*}[!t]
\centering
\includegraphics[width=1.0\textwidth]{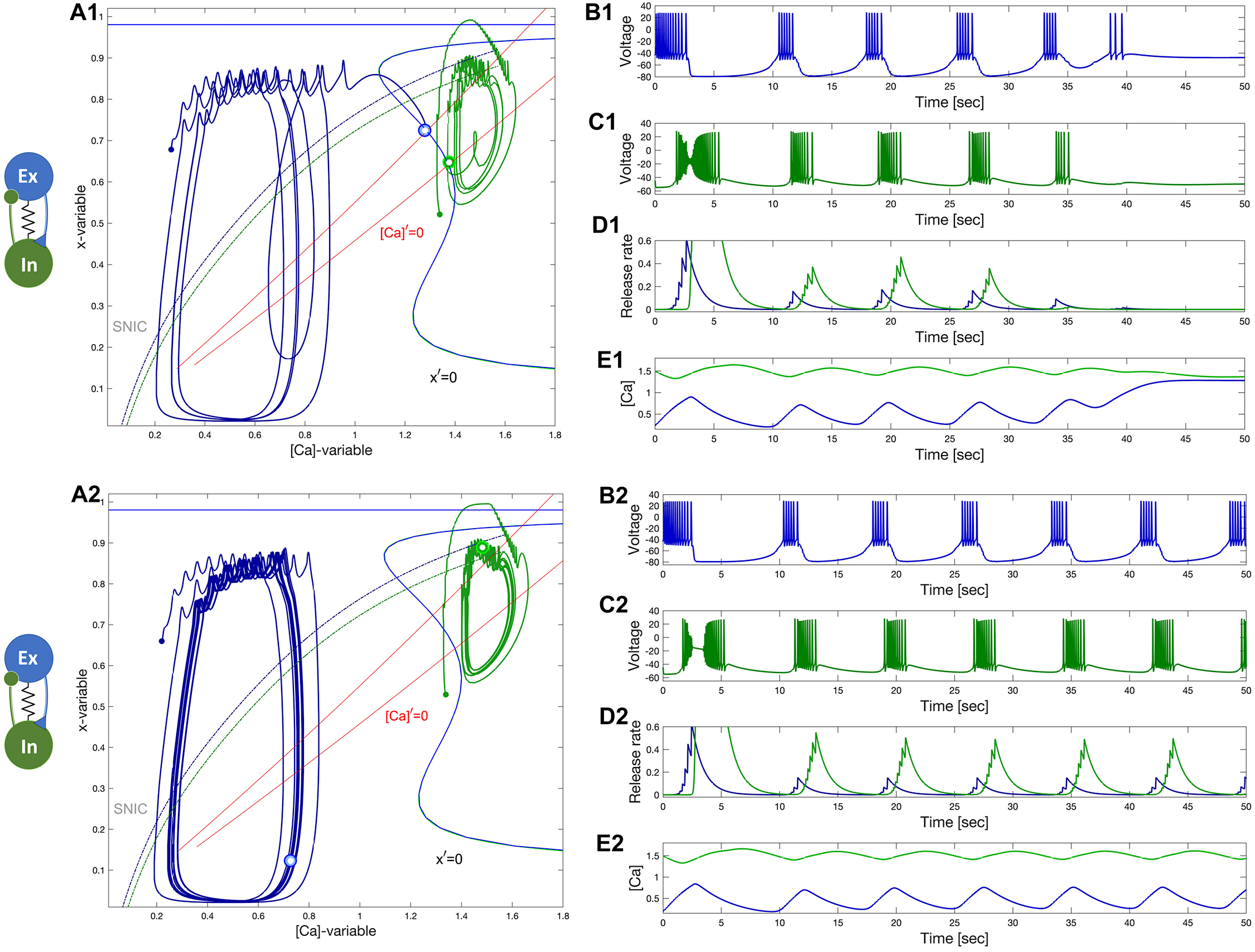}
\caption{Panels A1-E1:  Network bursting in the EI-module loses its initial momentum and transitions back to the native quiescent states: the excitatory (blue) neuron at $\Delta_{\rm Ca}=20$mV and the inhibitory (green) neuron  at $\Delta_{\rm Ca}=60$mV. (A1) Equilibrium states shown as blue and green double dots at the intersection of the nullclines $[\rm Ca]^\prime=0$  and $x^\prime=0$ in the $([\rm Ca],x)$-phase plane. Panels A2-E2: Increasing the forward excitation warrants self-sustained bursting oscillations with 1/4 phase-lag. Here $g_{elec}=2.5\times 10^{-4}$nS, $\alpha_{ex}=0.033$, $\beta_{ex}=0.001$, $\alpha_{in}=0.021$, $\beta_{in}=0.001$, $g_{ex}=0.0375$ and $0.045$, $g_{in}=0.6$}
\label{fig:fredo4}
\end{figure*}

Let us next elucidate the assembly process of the EI-module step by step. The assembly starts with the uncoupled neurons: blue tonic-spiking set at $\Delta_{\rm Ca}=-50$mV and green hyperpolarized quiescent set at $\Delta_{\rm Ca}=-20$mV, see  the voltage traces Fig.~\ref{fig:fredo1}B1 and C1. The location of the corresponding periodic orbit and stable equilibrium state in the $([\rm Ca],x)$-phase plane is determined by the corresponding two nullclines $[\rm Ca]^\prime=0$ (red) and the equilibrium nullcline $x^\prime=0$ in Fig.~\ref{fig:fredo1}A1. Next, after having the excitatory synapse installed, its positive drive (blue $S(t)$ in ~\ref{fig:fredo1}C1) shifts the inhibitory (green) neuron in the tonic-spiking mode as well, see Figs.~\ref{fig:fredo1}A2. Observe from figure~\ref{fig:fredo1}A2 that  the excitatory current  shifts  the equilibrium nullcline, $x^\prime_{\rm exc}=0$, rightwards relative to the position of and the intact nullcline $x^\prime=0$ (color-matching the neurons), in the $[\rm Ca]$-direction, respectively, as expected; compare with Figs.~\ref{fig:cascade}E and F above.

On the next step we gradually introduce a negative, inhibitory feedback loop, shown in Fig.~\ref{fig:fredo2}. Specifically, from Fig.~\ref{fig:fredo2}D1 one can observe the onset of network bursting emerging initially due to the strong reciprocal EI interaction  documented in the time progressions of the neurotransmitter rates $S(t)$. Nevertheless, as the network coupling through the positive and negative feedback loops is yet unbalanced, so the oscillations cannot self-sustain and seize back to spiking activity in both neurons. 
	
Increasing the inhibition balances out the nonlinear interactions through both feedback loops, so that the network starts generating slow bursting activity where the driven inhibitory neurons follow the driving excitatory neuron with the desired phase-lag around a quarter of the burst period. Introducing additionally a gap junction between the neurons can seize the emergent oscillations, unless they are well  self-sustained and sufficiently robust. In the latter case, the gap junction aids in equating the burst durations and hence the numbers of spikes per burst in both neurons (Figs.~\ref{fig:fredo2}B1 and C1) of the EI-module. This can be treated as an implicit indication of the structural stability of the network, namely its dynamics persist with adequately balanced parameter variations. Observe from the phase plane in Fig.~\ref{fig:fredo2}A2 that the inhibitory drive sends the excitatory (blue) into hyperpolarized quiescent deeps following the approximate position of the forced equilibrium nullcline $x'_{inh}=0$.  Note that the inhibitory feedback can become stronger with increasing the excitatory drive. The stronger the inhibition becomes, the longer the excitatory (blue) neuron recovers from its forced hyper-polarizing episodes. This imply that stronger coupling can prolong the interburst interval and hence burst period of the network. Meanwhile, increasing the strength of the bi-direction electric synapse (gap junction) lets the excitatory neuron recover from its hyperpolarized excursion faster by counter-acting the inhibition action, and therefore restore the network balance and bring the duty cycle of networking bursting closer to 1/2. Recall that the duty cycle is a ratio of the burst duration relative to burst period. However, meanwhile, bi-directionality can also affect the inhibitory neuron by shortening its active phase and bringing the duty cycle of network bursting close to the sought value 1/2 as well. This observation provides the means to quantify some balanced, stability conditions for the network self-sustained oscillations: i) equal burst durations in both neurons and ii) the proximity of the phase lag between them to the desired 1/4 of the burst period, which is clearly demonstrated in the $S(t)$-traces in Figs.~\ref{fig:fredo2}D2. 

\begin{figure*}[!t]
\centering
\includegraphics[width=1.0\textwidth]{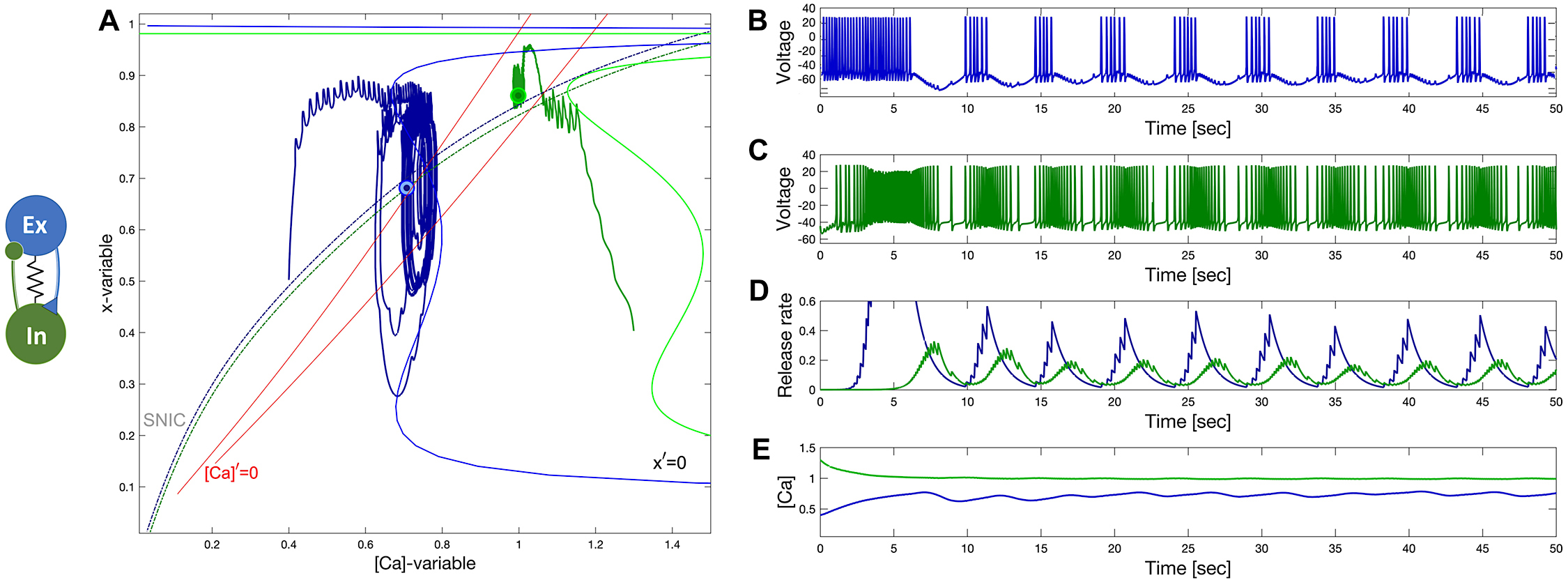}
\caption{Inhibition-driven bursting in the excitatory (blue) neuron  reciprocally causing a slow frequency modulation in the voltage trace of the inhibitory (green) neuron in the EI-module;  here both neurons are set to spike tonically, resp., at $\Delta_{\rm Ca}=-60$mV and  $\Delta_{\rm Ca}=-40$mV; $g_{elec}=4 \cdot 10^{-3}$S, $\alpha_{ex}=0.025$, $\beta_{ex}=0.001$, $\alpha_{in}=0.01$, $\beta_{in}=0.0014$, $g_{ex}=0.01$ and $g_{in}=0.15$ }
\label{fig:fredo5}
\end{figure*}
       
Note that bursting oscillations in the voltage trace of the inhibitory neuron occur due to its cycling between a meta-stable state due to excitation drive and its natural resting state, which it receives no drive while the excitatory neurons transitions throughout forced hyperpolarized episodes. These cycling translates into small oscillations crossing the SNIC-curve in the $([\rm Ca],x)$-plane around $[\rm Ca]=1.15$ in Figs.~\ref{fig:fredo2}A2 and E2. Unlike large-amplitude network limit cycle of the excitatory neuron, the oscillations corresponding to the inhibitory one are hardly noticeable. Nevertheless, as long as the transient stays above or below the SNIC-curve, the corresponding voltage traces will respond, respectively, with spike trains alternating with quiescent intervals, see  Figs.~\ref{fig:fredo2}B2 and C2.    

To demonstrate well-pronounced network bursting, its stability and parameter range, we make the neurons quite diverse: the excitatory neuron is set to be actively tonic-spiking at $\Delta_{\rm Ca}=-60$mV, while the inhibitory one is now deeply hyperpolarized at $\Delta_{\rm Ca}=20$mV, and follow the same assembly stages as above. The ultimate result is presented in Fig.~~\ref{fig:fredo3}. One can see from corresponding (green) cycle in the $([\rm Ca],x)$-phase plane in figure~~\ref{fig:fredo3}A that the inhibitory neuron now makes longer and deeper excursions between the two phases: i) tonic-spiking and ii) quiescent separated by the corresponding (green) SNIC-curve. The difference can be also observed in the voltage traces (Fig.~~\ref{fig:fredo3}B1 and B2), which leads to the longer network period of emergent bursting. In addition, Fig.~\ref{fig:fredo3}C demonstrates a stable limit cycle in the $\left ([\rm Ca]_1, [\rm Ca]_2 \right )$-phase plane which is an indicator of the proper balance of amplitudes and phases. While the coupling balance of inhibition and excitation results in the phase-lag close to the desired quarter (1/4) period. This is well documented in the $([\rm Ca],x)$- phase plane showing an instantaneous snapshot of the current phases of both neurons as two double dots: blue at 12 and green around 9 o'clock on the corresponding limit cycles. Lastly, we add that increasing the gap junction strength makes the temporal characteristics of bursting oscillations even more equal in both constituent neurons.

\subsection{Two quiescent neurons also lead to perpetual bursting motion}

Next, let us discuss two more options to generate emergent bursting in the EI-module. In the first case, both neurons are naturally quiescent at different $\Delta_{\rm Ca}$-values, as can be deduced from their voltage traces in Figs.~\ref{fig:fredo4}B1 and C1. This figure depicts that the excitatory neuron is initiated in the tonic-spiking phase far from its natural equilibrium (represented by the small blue dot) to produce a flux of positive feedback to the inhibitory neuron, which is initially set closer to its native equilibrium state (green dot). We can deduct from this figure that this EI-module is misbalanced and cannot keep up its initial momentum, as seen in the synaptic rate $S(t)$ traces in Fig.~\ref{fig:fredo4}D1, which decay and slow down emerging bursting back to the quiescent states in both neurons. Those are represented by the blue and green double-dots through which the (red) nullclines $[\rm Ca]^\prime=0$, left for $\Delta_{\rm Ca}=20$mV and right  for  $\Delta_{\rm Ca}=60$mV, cross the $\Sigma$-shape nullcline $x^\prime=0$ in the $([\rm Ca],x)$-projection in Fig.~\ref{fig:fredo4}A1.

Increasing the excitation remedies the situation, as one see from see Figs.~\ref{fig:fredo4}A2-E2. So, it appears that newly born stable limits cycles underwent through a network version of an Andronov-Hopf bifurcation (Fig.~\ref{fig:fredo4} A2) that gave rise to the onset of steady self-sustained bursting. This bursting is due to a strong or balanced level of reciprocal nonlinear interplay of excitatory and and inhibitory currents flowing back and forth through feedback looks between the constituent neurons.  Recall that while coupling can be sufficient for the occurrence of self-sustained bursting, it can fail at the initial stage because of an improper balance of the  phases or a phase-lag between the neurons in the EI-module. On the other hand, even with proper initial phases, the network may not gain the initial momentum, and bursting will slow down and seize just like in the previous case (Fig.~\ref{fig:fredo4}A1). This is an indicator that likewise the HCO, the EI-module is bi-stable as well, and moreover network oscillations are composed of transient trajectories generated by its components. This is reminiscent to the concept and principles underlying perpetual motions observed in various pairs of diversely connected mechanical bodies \cite{pmotion}.

Let us reiterate that the phase-lag between oscillations generated by the advancing excitatory neuron and the following inhibitory neuron is to be close to quarter of the network period. This is apparent in Fig.~\ref{fig:fredo4}A2 (as well as panel D2), which captures a snapshot of the relative positions of the blue and green phase points on the corresponding cycles at 5 (lower right) and 12 o'clock (above) respectively. Recall that the change rate of the gating $x$-variable is faster than that of $[\rm Ca]$-dynamics, and therefore the phase points do not turn along the cycles uniformly (with non-constant orbital speeds). Therefore, such snapshots can show some variability of the phase lags between oscillations fluctuating around the target value of 1/4 on average.

\subsection{Two tonic-spiking neurons in the excitatory-inhibitory module}

\begin{figure*}[!t]
\centering
\includegraphics[width=0.45\textwidth]{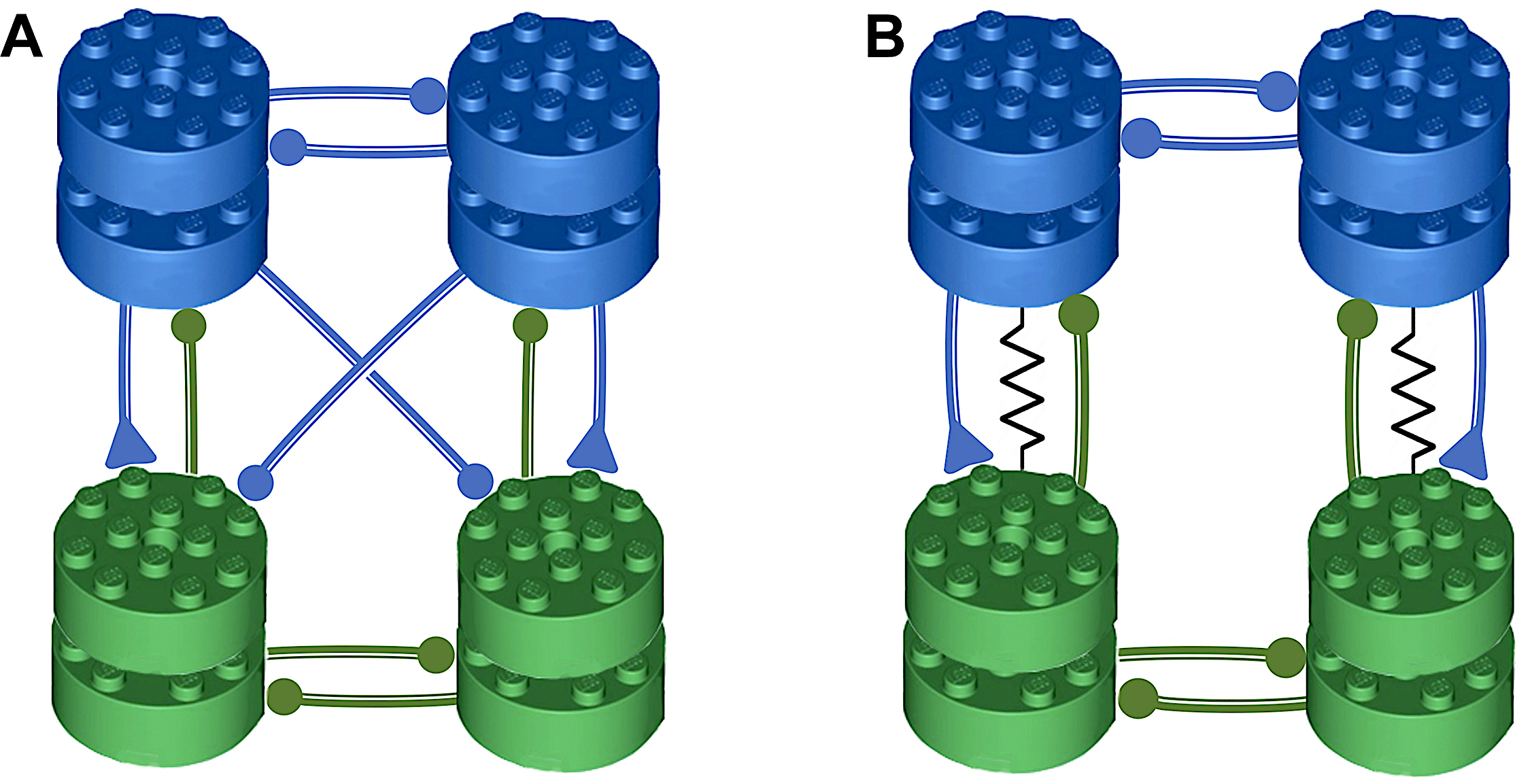}
\caption{Lego-style circuitries of the swim CPGs in the sea slugs {\em Melibe leonina} in Panel~A and {\em Dendronotus iris} in Panel~B, resp., which are assembled with the HCO and EI-module as their building blocks including  built-in feedback loops in them to promote and maintain robust and structurally stable bursting at the network-level. These are the untwisted representations compared to the conventional wiring circuitries of the biological sea slug CPGs.}
\label{lego}
\end{figure*}

The bistability and the conditions of balanced initial phases and mixed coupling remain valid also in the case of the excitatory-inhibitory module comprised of two tonically spiking neurons at two distinct values $\Delta_{\rm Ca}=-60$mV  and $\Delta_{\rm Ca}=-40$mV, see Fig.~\ref{fig:fredo5}. One can see from the voltage traces (Fig.~\ref{fig:fredo5}B and C) that the blue excitatory neuron demonstrates bursting activity with quiescent phases caused by the flux of the inhibitory current generated by the green inhibitory one. The stronger the inhibition current is, the longer the excitatory neuron recovers from it. This feedback mechanism alone can regulate the duty cycle of network bursting in the EI-module. Note that the inhibition becomes stronger as the excitatory drive increases; the inhibitory neuron receives the excitatory drive and becomes even more {\em hyper} active with a greater spike frequency through the forward loop and vice versa. Because both neurons are initially chosen to demonstrate tonic-spiking activity, the voltage trace of the inhibitory one in Fig.~\ref{fig:fredo5}C demonstrates clearly a spike frequency modulation caused periodically by the positive drive originating from the excitatory neuron. In turn, those cycling episodes of higher frequency cause stronger inhibition feedback looped back onto the excitatory neuron that warrants the quiescent interburst intervals in its traces, and so forth. At this point, the role of stronger electrical coupling between both neurons comes into play. As we mentioned earlier, the electrical gap junction equates the dynamics of the coupled neurons, and even synchronizes them in the absence of mixed chemical synapses. In the given case, depending on the inhibitory/excitatory balance, increasing the electric coupling can produce two possible outcomes: i) depolarizing the excitatory neuron and shortening or illuminating interburst intervals and bringing it back to original tonic-spiking activity. In the second case ii) inhibition is ``prevails'' over excitation, so that the interburst episodes when the excitatory neuron becomes deeply hyperpolarized around  $-70$ to$-75$mV, they also bring down the voltage in the inhibitory neuron,  and its spike frequency decreases substantially. One can observe fast EPSPs between bursts in the voltage trace (Fig.~\ref{fig:fredo5}B) of the excitatory neuron caused by the electrical current in correlated response to spikes generated by the inhibitory one. This ``equating'' action by the gap junction can be also seen in the position of the corresponding cycles in the $([\rm Ca],x)$-projection of Fig.~\ref{fig:fredo5}A: weakening the strength of electrical coupling noticeably increases the amplitude of forced oscillations in the blue excitatory neuron. On the contrary, gradual increasing the electrical coupling will misbalance the excitatory/inhibitory ratio, eventually bring both cycles closer in the phase plane, and narrowing the phase-lag between them will result in halting the burst rhythmogenesis in the excitatory-inhibitory module.

\section{Conclusion} \label{sec:7}

We laid out how to create slow rhythmic networks composed of the model neural adapted from the Plant burster, and the newly developed logistic model of slow synapses. We dissected the dynamics of the neuron model including two additional bifurcation parameters introduced in its slow subsystem which allows us to toggle the model between bursting, tonic-spiking, and quiescent neuron activity. The proposed model of the logistic synapse was shown to reproduce the experimentally observed nonlinear dependence of the synaptic current on the spike frequency in the biological swim CPG interneurons. This in practice allows the synapse to delay or act as a high-pass filter of the presynaptic neuron. The overall quality of the logistic synapse model is its plastic kinetics, which provides more flexibility and less stiffness and selectivity to parameter variations of network coupling, compared to other models. 

With our understanding of the Plant model and the creation of the experimentally plausible logistic synapse, we then began exploring the dynamic properties of two basic building blocks observed in various biological systems, mutually inhibited (HCO) and excitatory-inhibitory pairs of neurons. We argued that having modified the neuron model and its latent inability to generate bursting activity, could explore in detail the transient dynamics of the emergent bursting network. This exploration led to the novel concept of the network hysteresis, which happens due to the occurrence of cycling overlaps between the tonic-spiking and perturbed quiescent manifolds in the phase space of the model. In our exploration of the two configurations, we discussed various cases including those proceeding the stable before rhythm generation, as well as transient rhythm generation. 

Figure~\ref{lego} represents schematically the untwisted circuitries of the swim CPGs in the sea slugs {\em Melibe leonina} in Panel~A and {\em Dendronotus iris} in Panel~B, respectively. One can observe from this figure that both circuits are pair-wised assembled using the HCO and EI-module as their building blocks. This supports our hypothesis the all assembly parts in the given CPGs are designed to promote and maintain stably slow rhythm generation at the network level which is structurally stable to cellular and synaptic parameters, as well as robust and resistant to perturbations, intrinsic and external. This may seem as a surprising discovery due to overwhelmingly redundant design including several built-in feedback loops in the swim the CPGs because the locomotion of both sea slugs eventually results in a simple bilateral bending swimming. Nevertheless, the long evolutionary history of the sea slugs with fluctuating parameters yet capable of swimming in the broad range from 1 to 14 seconds during the life span for millions of generations is {\em per se} a de-facto proof of the success of the design solutions developed by Nature. 
	
We hope that this work can be used as an instructional guide to create diverse biologically plausible neuronal networks that produce various rhythms with various time scales based upon transient properties of nonlinear interactions among coupled interneurons.

\section*{Acknowledgments} \label{sec:8}
We are grateful to J.~Rinzel for historic insights, and R.~Poh for a careful proofreading of our manuscript. We thank the Brains and Behavior initiative of Georgia State University for the pilot grant support, as well as for the B\&B graduate fellowships awarded to J.~Bourahmah and J.~Scully, and a summer B\&B scholarship to F.~Padilla.  The authors thanks the current and former Shilnikov NeurDS  (Neuro-Dynamical Systems) lab-mates, specifically, F. Padilla, C.~Hinsley,  Drs. H.~Ju, D.~Alacam, and K.~Pusuluri for inspiring discussions, as well as  A.~Sakurai and P.S.~Katz for sharing their neurophysiological insights and knowledge with us.  This paper was partially funded by our joint National Science Foundation award IOS-1455527. 

\section*{Data availability}

The toolkit that supports the findings of this study is openly available and deposited in GitHub at \url{https://github.com/jamesjscully/plant_paper.} Our software codes including equations, parameters etc. are available upon request.

   \label{fig:supp}
\section*{Appendix I: a basic Hodgkin-Huxley based model of the swim CPG interneuron} \label{sec:appendix1}

The dynamics of the membrane potential, $V$, is governed by the following equation:
\begin{equation}
 	C_{m} {V}^\prime = -I_{I} - I_K - I_{T}  - I_{KCa} -  I_{h}  - I_{leak}  - I_{syn}.  \label{a1}
 \end{equation}
 The fast inward sodium and calcium $I_{I}$-current is given by   
\begin{equation}
I_{I}=g_{I}\,h\, m^{3}_{\infty}(V)(V-E_{I}), 
 \end{equation}
with the reversal potential $E_{I}=30$mV and and the maximal conductance value $g_{I}=4$nS, and  
\begin{align}
 m_{\infty}(V) &= \frac{\alpha_{m}(V)}{\alpha_{m}(V) +\beta_{m}(V)}, \qquad  \\
 \alpha_{m}(V) &= 0.1 \frac{50-V_s}{1+e^{(50-V_s)/10}}\ , \\ 
 \quad \beta_{m}(V) &= 4 e^{(25-V_s)/18}  
\end{align}
 the dynamics of its inactivation  gating variable $h$ is given by      
\begin{align}
 {h}^\prime &=  \frac{h_{\infty}(V)-h}{ {\tau_{h}(V)} }, \quad \mbox{where} \quad \\
h_{\infty}(V) &= \frac{\alpha_{h}(V)}{\alpha_{h}(V) +\beta_{h}(V)}\quad \mbox{and} \\
 \quad \tau_{h}(V) &= \frac{12.5}{\alpha_{h}(V) +\beta_{h}(V)} 
\end{align}
 \begin{align}
 \alpha_{h}(V) &= 0.07 e^{(25-V_s)/20}, \\
 \beta_{h}(V) &= \frac{1}{1+e^{(55-V_s)/10}}, \quad \mbox{and} \quad \\
 V_s &= \frac{127V+8265}{105}{\rm mV}.
 \end{align} 
  The fast potassium $I_K$-current is given by the equation  
\begin{equation}
I_K=g_{K} n^{4}(V-E_{K}),  
\end{equation}
with the reversal potential is $E_{K}=-75$mV and the maximal conductance is set as $g_K=0.3$nS. The dynamics of inactivation gating variable is described by 
\begin{align}
{n}^\prime &= \frac{n_{\infty}(V)-n}{\tau_{n}(V)}, \quad \mbox{with} \quad \\
n_{\infty}(V) &= \frac{\alpha_{n}(V)}{\alpha_{n}(V) +\beta_{n}(V)} \quad \mbox{and} \quad \\
\tau_{n}(V) &= \frac{12.5}{\alpha_{n}(V) +\beta_{n}(V)}, \quad \mbox{where} \quad \\
\alpha_{n}(V) &= 0.01 \frac{55-V_s}{e^{(55-V_s)/10}-1} \quad \mbox{} \quad \\
\beta_{n}(V) &= 0.125 e^{(45-V_s)/80}.
\end{align}	
The rapidly depolarizing h-current is given by 
\begin{equation}
I_{h}=g_{h} \frac{y (V-E_{h})}{(1+e^{-(V-63)/7.8)})^3}, 
\end{equation}
with $E_{h}=70$mV, and $g_h=0.0006$nS; the dynamics of its $y$-activation probability is described by 
 \begin{equation}
{y}^\prime = \frac{1}{2} \left [ \frac{1}{1+e^{10(V-50)}}-y \right ]/ \left [ 7.1+\frac{10.4}{1+e^{(V+68)/2.2}} \right ], 
\end{equation}	
whereas the leak current is given by 
\begin{align}
I_{leak} &=g_{L} (V-E_{L}), 
\end{align}
with $E_{L} =-40{\rm mV}$, and $g_{L}=0.0003{\rm nS}$. The sub-group of slow currents in the model includes the TTX-resistant sodium and calcium $I_{T}$-current given by   \begin{align}
 I_{T} &= g_{T} x (V-E_{I}),
\end{align}
with $E_{I} =30{\rm mV}$ and $g_{T}=0.01{\rm nS}$, where the dynamics of its slow activation variable are described by  
 \begin{align}
{x}^\prime &= \frac{x_{\infty}(V)-x}{\tau_{x}}, \quad \mbox{where} \\ 
\quad x_{\infty} &= \frac{1}{1+e^{-0.15(V+50-\Delta_x)}},
\end{align}\label{eqttx} 
with the time constant $\tau_{x}$ set as 100 or 235ms in this study. The slowest outward $Ca^{2+}$ activated $K^+$ current given by  
\begin{equation}
I_{KCa} =g_{KCa}\frac{[Ca]_i}{0.5+[Ca]_i}(V-E_{K})
 \label{IKCa}
\end{equation}
with  $E_{K}=-75\mbox{mV}$, and $g_{KCa}=0.03$, where the dynamics of intracellular calcium concentration are governed by 
\begin{equation}
{[Ca]}^\prime  =  \rho \left ( K_{c}\, x\, (E_{Ca}-V + \Delta_{Ca})-[Ca] \right  )  \label{eqCa}
\end{equation}
with the Nernst reversal potential  $E_{Ca}=140$mV, and small constants $\rho= 0.00015{\rm ms}^{-1}$ and  $K_c=0.00425{\rm mV}^{-1}$.

\section*{Appendix II: Implementation details for fast-slow decomposition} \label{sec:appendix2}

In this appendix we describe the computational processes used to determine the relative positions of the slow nullclines, SNIC bifurcation curves, and average quantities needed to locate periodic orbits,  as well as the heat maps used in the figures of the main text. Three approaches were used throughout the text, both separately and in combination.

\subsection{Approach I: Equations of slow nullclines} \label{sec:appendix2a}

To locate the positions of the slow nullclines $x'=0$ and $[\rm Ca]'=0$ in the $([\rm Ca]m\,x)$-phase plane
in the figures above, we first parametrize the membrane voltage $V_n$ within an equilibrium range $[-70;\,+20]$mV with some step size. Next, solve the equilibrium state equation $V'=0$ for the $I_{KCa}$-current:  
$$
 I_{KCa} = -I_{I} - I_K - I_{T}   -  I_{h}  - I_{leak}  - I_{syn},  
$$ 
with the corresponding static functions $h_\infty(V_n)$, $n_\infty(V_n)$, and $m_\infty(V_n)$ for equilibria in the currents above, and next using Eq.~(\ref{IKCa}) find the parameterized array   
$$
[{\rm Ca} (V_n)]= 0.5\, \frac{I_{KCa}}{g_{KCa} (V_n-E_k)-I_{KCa} }.   
$$ 
The ordered pairs $ \left\{  [{\rm Ca}(V_n)], \, x_\infty (V_n) \right \} $ populate the sought $x$-nullcline $x'=0$.     
To determine the position of the nullcline $[\rm Ca]'=0$,  first solve the equation $V'=0$ for the $I_{T}$-current: 
$$
I_{T} = -I_{I} - I_K -  I_{KCa}  -  I_{h}  - I_{leak}  - I_{syn},  
$$ 
next find 
$$
x(V_n)=\frac{I_{T}}{g_T (E_{I}-V_N)},
$$
and then from Eq.~(\ref{eqCa}) find 
$$
[{\rm Ca}(V_n)]=g_{T}\,x(V_n)\,(E_{Ca}-V_n+\Delta_{\rm Ca}),  
$$
and use these ordered pairs $\left \{ [{\rm Ca}(V_n)],\, x_\infty (V_n) \right \} $ to locate the  calcium nullcline $[\rm Ca]'=0$ in the slow $([\rm Ca]m\,x)$-phase plane.

\subsection{Approach II: Averaging approach to identify periodic orbits} \label{sec:appendix2b}

To appraise and quantify the behavior of the model during episodes of fast spiking activity due to its fast subsystem requires, we need to determine the location of a corresponding stable orbit on the $M_{\rm po}$-manifold and in its 2D pseudo-projection, as depicted in Figs.~\ref{fig:Plant_burster}A and B, respectively. To achieve this rigorously following Refs.\cite{Shilnikov2005,Mmo2005,Shilnikov-12}, one has to numerically solve the following averaged system adopted from Eqs.~(\ref{SlowSub1})-(\ref{SlowSub2}) for a pair $(\langle x \rangle^*, \langle [\rm Ca] \rangle^*)$:
 \begin{align}\label{aver1} 
\langle x \rangle^* &= \frac{1}{T} \int_{0}^T \frac{dt}{1+e^{-0.15(V_{\rm po}(t)+50- \Delta_{x})}},\\   
\langle [{\rm Ca}] \rangle^*  &=  K_{c}E_{Ca} \langle x \rangle^ + \Delta_{Ca} - \frac{ K_{c} }{T} \int_{0}^T x_{\rm po}(t)\,V_{\rm po}(t) dt ,
\end{align}
 where $V_{\rm po}(t)$,  $x_{\rm po}(t)$  and $[\rm Ca_{\rm po}(t)]$ are the coordinates of the periodic orbit with period $T$, situated on the critical 2D $M_{\rm po}$-manifold at the given  $(\Delta_{\rm Ca},\,\Delta_x)$-parameter values. Here $\langle x \rangle$ is the average quantity from Eq.~(\ref{aver}). The solution $(\langle x \rangle^*, \langle [\rm Ca] \rangle^*)$ of the average equation above corresponds to the periodic orbit with its  conceptualized  ``gravity center''.

In practice, there is a (computationally) cheap trick for finding a loose approximation for this average in this neuron model. By setting $\tau_x=1000$ or larger one makes the $x$-dynamics slow so that it cannot  longer echoes  fast voltage oscillations in numerical simulations of transient trajectories, but steadily traces out the sought average $\langle x \rangle$ nullcline in the $([{\rm Ca}],x)$-phase plane.

\begin{figure*}[!t]
   \centering
   \includegraphics[width=.54\textwidth]{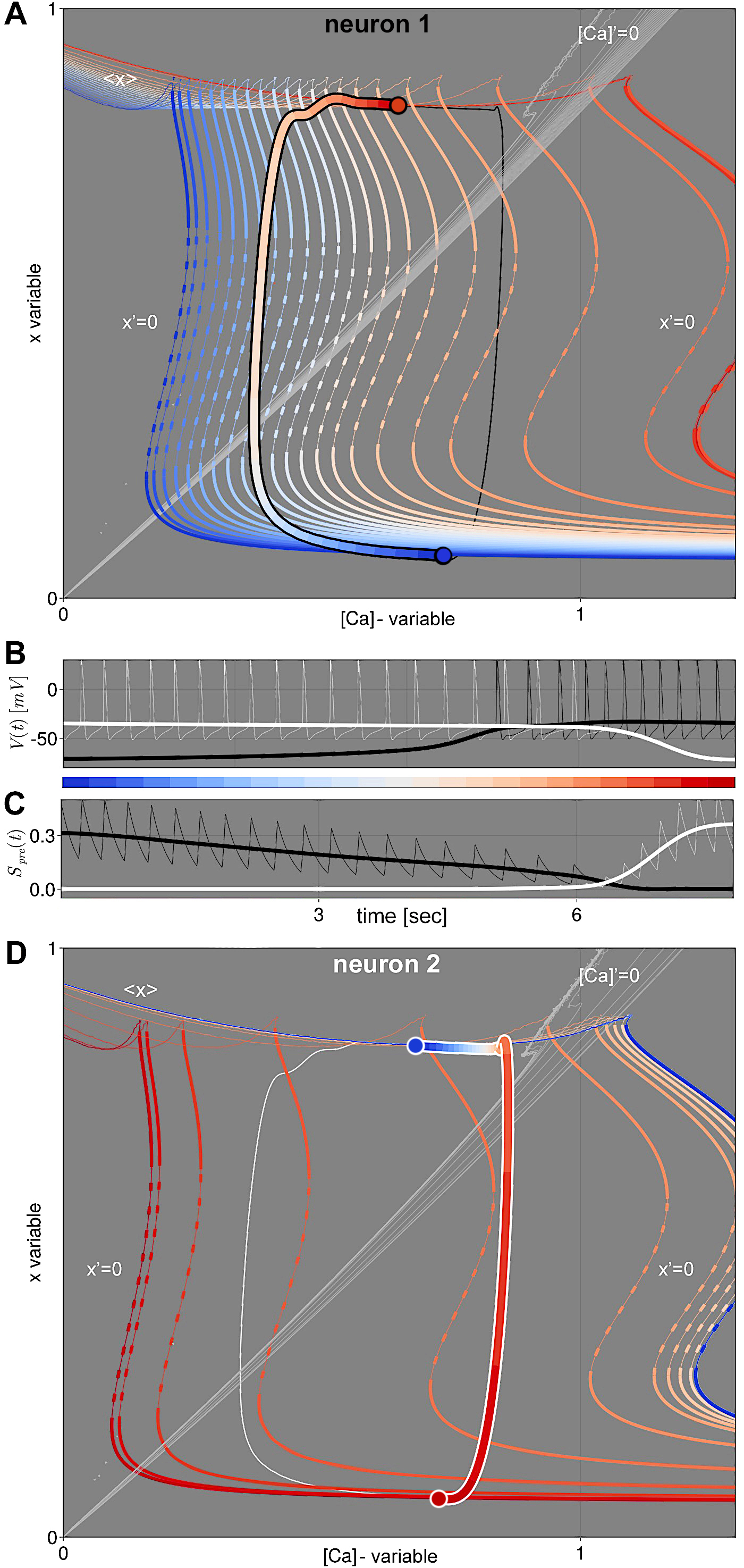}
   \caption{Illustration of the dynamic evolution of the slow nullclines, equilibrium $x'=0$, average $\langle x \rangle$, and calcium  $[\rm Ca]'=0$, in the slow $[\rm Ca,\,x]$-phase plane overlaid with the thick semicircular bursting orbits for both neurons in the HCO network over its half phase. Compare with Figs.~\ref{fig:hco1}B2-C2. In all panels the color spectrum, changing from blue through red, is used to communicate time progression. Black and white lines in all four panels are associated with neurons~1 and 2, respectively. The thick lines in Panel~B illustrate the smoothed the voltage traces $V(t)$ used to calculate the slow phase plane, whereas the thin lines illustrate the unaltered trajectory of the full system.  The time progressions of the synaptic variable $S_{pre}(t)$  are presented in Panel~C. Panel~A and D demonstrate slight variations of the calcium nullclines $[\rm Ca]'=0$, while the equilibrium nullclines $x'=0$ in the post-synaptic neurons are strongly affected by the reciprocal inhibition.}\label{fig:supp1}
   \end{figure*}

\subsection{Approach III: numerical scan and bi-parametric sweep} \label{sec:appendix2c}
    The second approach employs contour lines identified in bi-parametric sweeps of the fast subsystem, where the $x$ and $[\rm Ca]$ variables become ``frozen'' and treated as control parameters. First, we create a grid (of $[300]\times[300]$-resolution) of ordered $([\rm Ca],\,x)$-pairs over the slow phase plane. For each pair, a phase trajectory of the model computed and the initial transient is discarded. The trajectory is classified as either a spiker if voltage has three or more zero crossings, and quiescent otherwise.
    Each case is treated separately but in both cases, four quantities are characterized: spike frequency, and average values $\langle V \rangle $, $\langle x \rangle $ and $\langle [\rm Ca] \rangle $. When the system is quiescent, the frequency is 0, the
average voltage is the final voltage, which is simply
plugged into the $x$ and $[\rm Ca]$ state equations to determine $\langle x \rangle$ and $\langle [\rm Ca] \rangle$-values. 

	In the case of tonic-spiking activity, we use the final three zero crossings of the entire trajectory between, to determine the established the spike frequency and the average voltage.
The  $x$- and $[\rm Ca]$-coordinates are calculated for every point in the sampling of the limit cycle, and subsequently averaged. This order of operations is required due to nonlinearities in the model dynamics. The nullclines are plotted as the zero contours of the resulting $\langle x \rangle $ and $\langle [\rm Ca] \rangle$ arrays, while the SNIC-curve itself is evaluated as the zero contour of the frequency array. Average voltage is only used to generate the corresponding heat-map color-painting in the phase plane with a red hue for its oscillatory part above the SNIC-curve, and below it with a blue hue representing quiescent phases.    
 
\section*{Appendix III: Sequential phase plane analysis of network periodic orbits} \label{sec:appendix3}

In order to validate the mechanism underlying rhythmogenesis in the two network models we visualize the movement of the nullclines over the course of network bursting. We partitioned a network bursting trajectory into subsets where all $S(t)$ were monotonic in time, and visualized the transformations of the geometry as the network oscillations unfolded. During our investigation of the excitatory-inhibitory module, we noticed that it was possible to obtain network oscillations with frozen $x$ and $[{\rm Ca}]$-variables, suggesting that the mechanism of rhythmogenesis was not a product of the slow phase dynamics alone. Further investigation into the excitatory inhibitory module is required. 

  We showed above that in the case of the HCO network, the onset of network oscillations emerging from transients is coordinated by cyclicly moving $x$-nullclines $x'=0$ in the slow $([\rm Ca],\,x)$-phase plane. Here, we include a supplementary Fig.~\ref{fig:supp1} to further detail and  validate our analysis. Its panels~A and D represent two representative snapshots of time-varying rearrangements of the nullclines, equilibrium $x'=0$, average $\langle x \rangle$ and calcium $[\rm Ca]'=0$, in a chronological sequence over a half of each burst cycle. Since the HCO network is symmetric, the neurons will next switch their roles to complete the full bursting cycle. Close examination yields an interpretation by breaking a half-cycle sequence into two stages:\\
   Stage~I. As the active pre-synaptic  neuron~2 (its initial/final positions given by blue/red dots in Panel~D) drifts rightwards and approaches the SNIC-curve (not shown) along the average $\langle x \rangle$ nullcline, its spike frequency decreases so that the synaptic probability $S_2(t)$ (black line in Panel~C) drops to zero. The concurrent decay of inhibition projected from neuron~2 onto neuron~1 (its initial/final positions given by blue/red dots in Panel~A) lets the corresponding $x_1$-variable rise, which in turn makes the average $\langle x \rangle$ nullcline for neuron~2 shift to the right. \\
   Stage~II. After the $x_1$-variable reaches the average $langle x \rangle $ nullcline (color-coded red line in Panel~A), neuron~1 starts spiking and hence inhibiting neuron~2. The inhibition causes the average $\langle x \rangle$ nullcline for neuron~2 to shift leftwards so that the $x_2$ variable drop down close to zero and that the corresponding phase trajectory of neuron~2 is about to start traversing along the forced (red line) equilibrium nullcline $x'=0$ in the phase plane in Fig.~\ref{fig:supp1}D, and so forth.  
  
  To smooth the $V(t)$ and $s(t)$ traces of the active spiking neurons, the voltage trajectory is partitioned to ensure that each spike belongs to a unique subset by monitoring a threshold set around -35mV. A maximum spike time of $1$s determines whether each subset contains a spike, or a quiescent episode of the burst. Spike containing subsets are replaced with their voltage averages, while quiescent voltage level are left intact. The partitioning and averaging for each $S(t)$ follow their corresponding presynaptic voltages. Finally, a composition of two moving averages with $0.5$s window is applied as a filter for further  smothering.

\section*{References}
\bibliographystyle{unsrt}

\end{document}